\documentclass[manuscript,screen]{acmart}

\AtBeginDocument{%
  }

\setcopyright{acmlicensed}
\copyrightyear{2018}
\acmYear{2018}
\acmDOI{XXXXXXX.XXXXXXX}

\acmConference[Conference acronym 'XX]{Make sure to enter the correct
  conference title from your rights confirmation emai}{June 03--05,
  2018}{Woodstock, NY}

\acmISBN{978-1-4503-XXXX-X/18/06}

\usepackage{xcolor}
\usepackage{upgreek}
\usepackage{commath}
\usepackage{multicol}
\usepackage{multirow}
\usepackage{threeparttable}
\usepackage{bm}
\usepackage{pifont}

\usepackage{graphicx} 
\usepackage{array}    

\usepackage[most]{tcolorbox}
\usepackage{xcolor} 

\tcbset{
  tech-tip/.style={
    colframe=black,
    colback=yellow!5,         
    fonttitle=\bfseries,    
    boxrule=0.2mm
  },
  finding/.style={
    colframe=blue!30!black,
    colback=blue!5,
    coltitle=white,
    fonttitle=\bfseries,
    boxrule=0.5mm
  }
}

\newtheorem{definition}{Definition}

\begin{document}

\title{A Survey on Private Transformer Inference}
\author{Yang Li}
\email{yang048@@e.ntu.edu.sg}
\orcid{xxx}
\authornotemark[1]
\affiliation{%
  \institution{Nanyang Technological University}
  \country{Singapore}
}

\author{Xinyu Zhou}
\affiliation{%
  \institution{Nanyang Technological University}
  \country{Singapore}}
\email{xinyu003@e.ntu.edu.sg}

\author{Yitong Wang}
\affiliation{%
  \institution{Nanyang Technological University}
  \country{Singapore}}
\email{yitong002@e.ntu.edu.sg}

\author{Liangxin Qian}
\orcid{0000-0002-7686-4580}
\affiliation{%
  \institution{Nanyang Technological University}
  \country{Singapore}}
\email{qian0080@e.ntu.edu.sg}

\author{Jun Zhao}
\affiliation{%
  \institution{Nanyang Technological University}
  \country{Singapore}}
\email{junzhao@ntu.edu.sg}

\renewcommand{\shortauthors}{Yang et al.}

\begin{abstract}
  
\end{abstract}



\maketitle

\section{Introduction}
Transformer models have emerged as game-changers to 
revolutionize the field of Artificial Intelligence (AI). 
For instance, both ChatGPT~\cite{ChatGPT} and Bing~\cite{Bing} have made the power of transformer-based models widely accessible, democratizing advanced AI capabilities.
These models leverage attention mechanisms~\cite{vaswani2017attention} adeptly to capture long-range dependencies in sequences of input tokens, allowing them to accurately model contextual information. Besides, unlike traditional task-specific learning approaches, large transformer models (e.g., GPT~\cite{radford2019language} and BERT~\cite{devlin2018bert}) are trained on huge quantities of unlabeled textual data and are directly useful for a wide variety of applications such as sentiment analysis, language translation, content generation, and question answering. 

However, the application of large transformers still presents certain risks, particularly regarding privacy issues~\cite{lund2023chatting,tlili2023if}. 
Most popular transformer models operate in a pattern called Machine Learning as a Service (MLaaS), where a server provides the model and inference services to users who own the data.
For instance, 
OpenAI provides ChatGPT as an online platform and offers remote APIs for developers, allowing users to access services by submitting prompts or messages.
Nevertheless, this pattern raises privacy concerns: users need to transmit their private data to a company's server and have no direct control over how their data is handled. They must trust that the server processes the data honestly and follows the agreed terms of service. There exists a risk that the server could misuse the data, including unauthorized processing, storing the data indefinitely, or even selling it to third parties.
Even if the server is trustworthy, there are inherent risks associated with centralized data storage, such as data breaches or unauthorized access by malicious insiders. These risks are particularly concerning when dealing with sensitive or personally identifiable information. Therefore, while MLaaS offers significant convenience and computational power, it also necessitates careful consideration of privacy issues.
As a result, serious user data privacy concerns have been raised, which even led to a temporary ban of ChatGPT in Italy~\cite{mauran2023whoops,lomas2023italy}.
Hence, there exists a gap between high-performance transformer inference services and privacy concerns, motivating the study of \textit{Private Transformer Inference} (PTI).

Private inference is a cryptographic protocol that allows for model inference while ensuring that the server gains no knowledge about the users' input, and the users learn nothing about the server's model, apart from inference results. 

Recently, private inference on transformers has been achieved by using private outsourced computation techniques, such as secure Multi-Party Computation (MPC)~\cite{yao1982protocols} and Homomorphic Encryption (HE)~\cite{gentry2009fully}.
MPC enables multiple parties to jointly compute a function over their inputs while keeping those inputs private. Its essence is to allow the computation of results without any party revealing their private data to others.
HE provides a way to perform computations on encrypted data without needing to decrypt it. This means that data can be processed in its encrypted form, preserving privacy and security throughout the computation process. HE is extremely useful in MLaaS as it allows clients to encrypt their data before sending it to the cloud for processing. The server can then compute the model’s functions directly on the encrypted data, providing results without ever accessing the raw data.


In this paper, we review state-of-the-art papers that implement PTI using different private computation techniques, primarily focusing on MPC and HE technologies. We discuss their pros and cons and propose some future avenues to tackle. Furthermore, we propose a set of evaluation guidelines to compare solutions in terms of resource requirements.
The key contributions of this paper are summarized as follows:
\begin{enumerate}
    \item Provide a comprehensive literature review of proposed solutions in the field of secure transformer inference from recent three years (2022-2024);
    \item Provide a breakdown of the main challenges faced in this field, as well as typical solutions that are implemented to mitigate them;
\end{enumerate}

The rest of the survey is organized as follows: In Section~\ref{sec:background}, we introduce background information on the transformer architecture and cryptographic primitives. In Section~\ref{sec:PTI}, we provide an overview of the selected PTI studies, the approaches, strengths and weaknesses. 
Linear layers in transformers (i.e., large matrix multiplications) and the efficient computing protocols under HE mechanisms are discussed in Section~\ref{sec:linear}. Non-linear layers (e.g., Softmax, GELU) in transformers and their corresponding efficient protocols are discussed in Section~\ref{sec:non_linear_layers}. In Section~\ref{sec:evaluation}, we compare the experimental results presented by the selected papers and analyze them based on different metrics. In Section~\ref{sec:challenges}, we discuss challenges and future directions. Section~\ref{sec:conclusion} is the summary of the whole survey.




\section{Background}\label{sec:background}

\subsection{Private Inference}

Consider a scenario where a client $\mathcal{C}$ possesses private data $x$, and a server $\mathcal{S}$ hosts a model $\mathcal{M}$. The client $\mathcal{C}$ aims to utilize $\mathcal{M}$ to compute the inference result $\mathcal{M}(x)$. The privacy requirement is that the server learns nothing about $x$, and the client learns nothing about $\mathcal{M}$ except what can be inferred from the $\mathcal{M}(x)$. Following~\cite{zhang2024secure}, we formally define private inference as follows:
\begin{definition}
    A protocol $\Pi$ between server $\mathcal{S}$ with model $\mathcal{M}$ and client $\mathcal{C}$ with input $x$ is considered private if it satisfies:
    \begin{itemize}
        \item \textbf{Correctness.} The final output of protocol $\Pi$, denoted as $y$, is as same as the correct inference result $\mathcal{M}(x)$.
        \item \textbf{Security.}
        Here we introduce the concept of Simulator $(\mathrm{Sim})$ to define the protocol security. A simulator is a hypothetical algorithm that mimics the behavior of a party during protocol execution. 
        \begin{itemize}
            \item \textbf{Corrupted client.} Even if client $\mathcal{C}$ is corrupted, there still exists an efficient simulator $\mathrm{Sim}_{\mathcal{C}}$ such that $\mathrm{View}_{\mathcal{C}}^{\Pi} \approx \mathrm{Sim}_{\mathcal{C}}(x,y)$, where $\mathrm{View}_{\mathcal{C}}^{\Pi}$ denotes corrupted $\mathcal{C}$'s view during the execution of $\Pi$, i.e., an attacker on $\mathcal{C}$ cannot get more information through the execution than through the simulator.
            \item \textbf{Corrupted server.} Even if server $\mathcal{S}$ is corrupted, there still exists an efficient simulator $\mathrm{Sim}_{\mathcal{S}}$ such that $\mathrm{View}_{\mathcal{S}}^{\Pi} \approx \mathrm{Sim}_{\mathcal{S}}(\mathcal{M})$, where $\mathrm{View}_{\mathcal{S}}^{\Pi}$ denotes corrupted $\mathcal{S}$'s view during the execution of $\Pi$.
        \end{itemize}
    \end{itemize}
\end{definition}

This definition provides a strong guarantee of privacy, ensuring that the inference process does not compromise data and model confidentiality. Please note that private inference does not consider attacks based on inference results (e.g., model inversion), and the protection from such attacks falls outside the scope of this study.


\subsection{Transformer Architecture}
\begin{figure}[h]
  \centering
  \includegraphics[width=\linewidth]{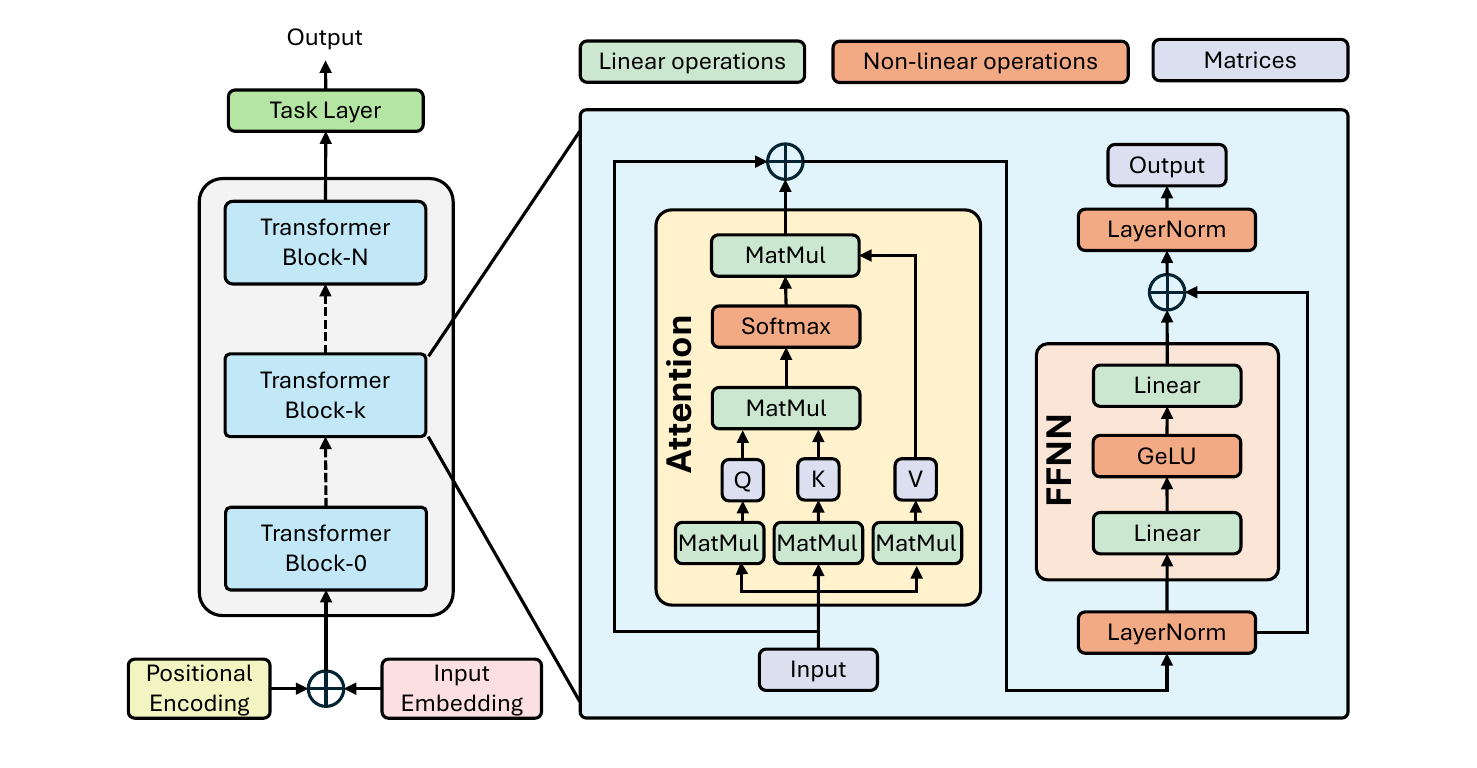}
  \caption{Structure and workflow of a Transformer~\cite{zhang2024secure}.}
  \Description{Transformer.}
  \label{Fig:transformer}
\end{figure}

Transformer is an encoder-decoder architecture in which both parts have a similar structure. We mainly focus on the encoder here.
The encoder comprises a stack of identical blocks, each with two sub-layers: a multi-head self-attention mechanism and a feed-forward network. Residual connection and layer normalization (LayerNorm) are utilized around each of the two sub-layers. An encoder's architecture and process flow are shown in Fig.~\ref{Fig:transformer}.
\\
\textbf{Embedding.} At the start of the encoder, an embedding layer is 
employed to transform input tokens into continuous feature vector representations. Concretely, given $\boldsymbol{X}_{\textnormal{input}}\in\mathbb{R}^{m\times1}$, an embedding lookup table is used to generate output $\boldsymbol{X}\in\mathbb{R}^{m\times d}$, where $m$ denotes the length of tokens and $d$ represents the model dimension.\\
\textbf{Attention.} Attention layers capture context and dependencies among the input through the multi-head attention mechanism. Specifically, an input $\boldsymbol{X}\in\mathbb{R}^{m \times d}$ is fed into $L$ multi-head attention layers. Each layer linearly projects the input $\boldsymbol{X}$ using the query, key and value weights $\boldsymbol{W}^Q_h,\boldsymbol{W}^K_h,\boldsymbol{W}^V_h\in\mathbb{R}^{d \times d'}$, where $h\in[H]$, $d'=d/H$ and $H$ is the number of heads in each multi-head attention layer. The results of the projections are $\boldsymbol{Q}_h,\boldsymbol{K}_h,\boldsymbol{V}_h
\in\mathbb{R}^{m \times d'}$.
Then, the attention mechanism works as the following function:
\begin{align}
    \mathrm{Attention}(\boldsymbol{Q}_h,\boldsymbol{K}_h,\boldsymbol{V}_h)= \mathrm{Softmax}\Big(\frac{\boldsymbol{Q}_h\boldsymbol{K}_h^{T}}{\sqrt{d}}\Big)\boldsymbol{V}_h,\label{fun:attention}
\end{align}
where the output can be denoted as $\boldsymbol{X}^{\textnormal{Att}}_h\in\mathbb{R}^{m\times d'}$ for simplification. Furthermore, all attention results $[\boldsymbol{X}^{\textnormal{Att}}_h\mid_{h\in[H]}]$ are concatenated to generate the final output $\boldsymbol{X}^{\textnormal{Att}}\in\mathbb{R}^{m\times d}$.\\
\textbf{Feed-Forward.} A fully connected feed-forward layer consists of two linear transformations with a Gaussian Error Linear Unit (GeLU) activation in between, which can be formalized as follows:
\begin{align}
    \mathrm{FeedForward}(\boldsymbol{X})= \mathrm{GeLU}(\boldsymbol{X}\boldsymbol{W}_1+\boldsymbol{b}_1)\boldsymbol{W}_2+\boldsymbol{b}_2.
\end{align}
\textbf{LayerNorm.} The LayerNorm is applied after the attention and feed-forward layers, ensuring that the inputs across different layers have a consistent mean and variance for stabilization. Given an input matrix $\boldsymbol{X}\in\mathbb{R}^{m \times d}$, we have:
\begin{align}
   \mathrm{LayerNorm}(\boldsymbol{X})_{i,j}=\frac{\gamma_j(x_{i,j}-\mu_i)}{\sigma_i}+\beta_j,
\end{align}
where $i\in[m]$, $j\in[d]$, $\mu=\sum_{i=1}^dx_i/d$ and $\sigma=\sqrt{\sum_{i=1}^d(x_i-\mu)^2}$ are mean and standard deviation, 
and $\boldsymbol{\gamma}$ and $\boldsymbol{\beta}$ are affine transform parameters.



The primary privacy concern arises because clients have no direct control over how their data is handled once it is sent to the server. They must trust that the server processes the data honestly and follows the agreed terms of service. However, there exists a risk that the server could misuse the data, including unauthorized processing, storing the data indefinitely, or even selling it to third parties.
Even if the server is trustworthy, there are inherent risks associated with centralized data storage, such as data breaches or unauthorized access by malicious insiders. These risks are particularly concerning when dealing with sensitive or personally identifiable information. Therefore, while MLaaS offers significant convenience and computational power, it also necessitates careful consideration of privacy issues.

\subsection{Privacy Preservation Techniques} \label{subsec:cryptography}
This subsection introduces the popular cryptographic techniques involved in current PTI studies.

\subsubsection{Functional Encryption} Functional encryption is a form of encryption that allows evaluating certain functions over encrypted data. The results of these functions are ``leaked'' from the ciphertexts, i.e., the result of the function is in plain data. This is beneficial for inference computation since only the first layer of the model needs to be run on encrypted data, and the rest can be run on plain data.
However, this does leak some client data information to the
server; the server learns the model output and the intermediate results of the computation.

\subsubsection{Secure Multi-Party Computation (MPC)}

Secure MPC~\cite{evans2018pragmatic} refers to a collection of cryptographic algorithms and protocols, e.g., garbled circuits, secret sharing, and oblivious transfer, which allows
multiple parties to jointly implement a computation task while keeping their respective data private. 
In essence, although all parties collaborate to perform a computation, no one can access information that compromises the privacy of others. 
The objective of MPC is to develop a secure protocol enabling these participants to collectively evaluate a function on their private inputs, ensuring that the output is accurate while safeguarding their private data, even in the presence of curious or malicious participants. Secure MPC can be formally described as follows:
Consider $n$ parties, denoted as $(\mathcal{P}_1,\ldots,\mathcal{P}_n)$, where each party $\mathcal{P}_i$ holds a private input $x_i$. The goal is to jointly compute a predefined function $f(x_1,\ldots,x_n) \rightarrow y$, where $y$ is the output obtained from the inputs of all parties. The computation results in $y=(y_1,\ldots,y_n)$, where each party $\mathcal{P}_i$ only learns $y$ (or a portion of $y$) without gaining any knowledge of the other parties’ inputs.

Upon these primitives, researchers have developed different protocols to implement model inference while preserving the privacy of the inputs. However, these protocols often require a lot of communication during the computation, which can make inference latency a problem. 

Furthermore, we introduce some of the basic building blocks in MPC. It is worth noting that we mainly present MPC techniques involved in the PTI studies, and the reader is advised to read~\cite{evans2018pragmatic,zhao2019secure} if interested in more MPC techniques.
\begin{itemize}
    \item \textbf{Secrete Sharing (SS):} Secret sharing is a fundamental component of many MPC techniques. Typically, a $(t,n)$-secrete sharing scheme divides a secret $S$ into $n$ shares so that any $t-1$ of the shares reveal no useful information about $S$. The secret can only be reconstructed from any combination of at least $t$ shares.
    \item \textbf{Oblivious Transfer (OT):} The oblivious transfer allows a sender to transmit one of many pieces of information to a receiver without knowing which piece was received. A standard 1-out-of-$k$ OT is denoted by $k\textnormal{-OT}$, where a sender $\mathcal{S}$ has $k$ messages $\{m_0,\ldots,m_{k-1}\}$ and a receiver $\mathcal{R}$ holds a choice bit $b\in[k]$. OT allows $\mathcal{R}$ to obtain $m_b$ while learning nothing about other messages, and $\mathcal{S}$ learns nothing. 
\end{itemize}

\subsubsection{Homomorphic Encryption (HE)}
HE is a form of encryption that enables computations on encrypted data directly without decryption. The computations remain encrypted and, once decrypted, yield results identical to those obtained from the same operations on unencrypted data. 
Most HE schemes are based on public key cryptography, where a public key $(\mathrm{pk})$ is used to encrypt data and a secret key $(\mathrm{sk})$ to decrypt results. The public key can be shared freely for encryption purposes, whereas the secret key is required to decrypt messages. These schemes are secure because access to the public key does not compromise the private key. An overview of the main homomorphic operations is as below.

\begin{itemize}
    \item $\mathrm{Enc}(\mathrm{pk},m)\rightarrow c$: On public key $\mathrm{pk}$ and a plaintext $m$, perform encryption to obtain a ciphertext $c$.
    \item $\mathrm{Dec}(\mathrm{sk},c)\rightarrow{m}$: On secret key $\mathrm{sk}$ and a ciphertext $c$, perform encryption to obtain the plaintext message $m$.
    \item $\mathrm{Eval}(\mathrm{pk},c_0,\ldots,c_{k-1},\mathrm{C})\rightarrow \mathrm{Enc}(\mathrm{pk},\mathrm{C}(m_0,\ldots,m_{k-1}))$: 
    On public key $\mathrm{pk}$, ciphertexts $c_1,\ldots,c_k$ encrypted from $m_1,\ldots,m_k$, and a circuit $\mathrm{C}$ (sequence of operations), outputs an encrypted computation result as same as the evaluation result of $\mathrm{Enc}(\mathrm{pk},\mathrm{C}(m_0,\ldots,m_{k-1}))$.
\end{itemize}

Nonetheless, every operation on encrypted data introduces a small amount of noise, which accumulates as more operations are performed. Beyond a certain threshold, this noise can become large enough to prevent correct decryption.
We can categorize HE schemes by the operations used in circuit $\mathrm{C}$ and its computational depth, i.e., the number of consecutive operations required for evaluation. 

\begin{itemize}
    \item \textbf{Partially Homomorphic Encryption:} Schemes support the evaluation of circuits consisting of a single type of operation—either addition or multiplication—but not both.
    \item \textbf{Somewhat Homomorphic Encryption (SHE):} SHE schemes enable a limited number of addition and multiplication operations on encrypted data, but this capability is restricted to a subset of computational circuits. The primary limitation arises from the accumulation of noise with each operation, which inherently restricts the depth of circuits that can be processed. 
    \item \textbf{Fully Homomorphic Encryption (FHE):} FHE schemes allow for arbitrary operations on encrypted data, unrestricted by the type or the number of operations. A key feature of FHE is \textit{bootstrapping}, a process designed to manage noise accumulation during operations. 
    Bootstrapping involves decrypting a ciphertext $c$ and then encrypting it again, leading to a fresh ciphertext $c'$ with a reset noise level. This could enable continuous computations on encrypted data. However, the computational intensity of bootstrapping makes FHE schemes still prohibitively expensive for many applications.
    \item \textbf{Leveled FHE.} Leveled FHE is designed to handle a predefined maximum number of operations on encrypted data. It supports both addition and multiplication, but requires the maximum circuit depth (number of operations) to be set in advance. Unlike standard FHE, Leveled FHE does not require the computationally expensive bootstrapping process. This allows it to provide an exact and adaptable depth of computation that can be precisely tailored for specific tasks, offering a balance between computational power and efficiency.
\end{itemize}






\section{Pivacy Threats in Secure Inference}

\textbf{Semi-honest.} The semi-honest security model is based on the assumption that all parties involved in computation will honestly follow the established protocols while passively attempting to gather extra private information during its execution. This model is often used when the parties have a basic level of trust in each other, not to actively disrupt the process. 
Semi-honest security is a common assumption for privacy-preserving machine learning (PPML).  The parties have to trust each other to some extent so that they can jointly contribute to the result of inference.

\textbf{Honest-Majority.}
The honest-majority security model operates under the assumption that as long as the majority of the participants are honest and follow the protocol, they can prevent a minority who may be dishonest from tampering with the process or stealing private information. 
It is suited for situations involving more participants, where not everyone might be fully trustworthy. This model enhances security by allowing the computation to tolerate some level of misbehavior from a minority of the participants without compromising the integrity or confidentiality of the overall process.


\section{An overview of private transformer inference studies: setups, cryptographic approaches, pros and cons}\label{sec:PTI}

This section provides an overview of state-of-the-art private transformer inference (PTI) studies, along with a critical review of their strengths and weaknesses. 
Specifically, we first introduce a secure inference system setup in Section~\ref{subsec:setup}. Then, we review current PTI studies and their privacy-preserving approaches in Section~\ref{subsec:overview}. A discussion about strengths and weaknesses is finally presented in Section~\ref{subsec:pros}.

\subsection{Bottlenecks in private transformer inference}

In recent years, there has been significant development in neural network private inference for traditional CNN (Convolution Neural Network) and RNN (Recurrent Neural Network) models~\cite{kumar2020cryptflow,tan2021cryptgpu}.
However, due to essentially different structures, private
transformer inference brings several new challenges.

\textbf{Large Matrix Multiplications.} First, transformer models involve a huge number of large matrix multiplications rather than matrix-vector multiplications widely studied in CNN. In addition, a transformer model usually consists of multiple encoders (i.e., decoders), and the multiplication scale is much higher than the traditional networks.
Direct extensions of existing matrix-vector multiplication protocols to transformers typically result in unaffordable overheads~\cite{hao2022iron,chen2024securetlm}.

\textbf{Complex Nonlinear Functions.} Traditional neural networks commonly employ crypto-friendly non-linear functions, e.g., Rectified Linear Unit (ReLU), Maxpool and batch normalization. In Contrast, transformers extensively employ Gaussian Error Linear Unit
(GeLU), Softmax and LayerNorm functions. These complex nonlinear functions are critical to the inference performance but are not computation-friendly for encryption primitives, thus making inference less efficient.

\subsection{An overview of PTI studies with employed cryptographies}\label{subsec:overview}


We examine the cutting-edge PTI studies (2022-2024) in Table~\ref{tab:overview}, including their setups, employed tools, threat models and improved components. In particular, the studies are classified and discussed based on their running setups~(2PC, 2PC-Dealer and 3PC). We believe that the choice of setups significantly affects the tools employed, the threat model, and the components to be improved. For instance, secure matrix multiplication requires additional privacy protection (e.g., HE) and overheads under the 2PC setup and is therefore optimized in all 2PC-based studies, but not in 2PC-Dealer and 3PC setups~\cite{li2022mpcformer,wang2022characterization,gupta2023sigma,luo2024secformer,chen2024securetlm,santos2024curl,dong2023puma,akimoto2023privformer,liu2024pptif}.

Studies based on 2PC~\cite{chen2022x,hao2022iron,hou2023ciphergpt,lu2023bumblebee,ding2023east,pang2024bolt,zhang2024secure} follow the most common server-client-participated computation paradigm and often assumes a standard semi-honest (i.e., 
honest-but-curious) threat model. The two parties, often server and client, will honestly follow the inference protocol while passively attempting to gather extra private information during its execution, which is a natural and practical consideration for current MLaaS platforms. In particular, THE-X~\cite{chen2022x} and NEXUS~\cite{zhang2024secure} manage to implement purely HE-based PTI frameworks, which requires them to optimize almost all nonlinear layers, as nonlinearity is often not supported by HE. In contrast, other studies~\cite{hao2022iron,hou2023ciphergpt,lu2023bumblebee,ding2023east,pang2024bolt} favor MPC+HE hybrid frameworks to improve the performance (e.g., inference speed, model accuracy) at the cost of additional complexity in the setups. As we have discussed, secure matrix multiplication in 2PC is the bottleneck for optimization as it requires additional privacy protection.

In both 2PC-Dealer and 3PC setups, the introduction of a "helper party" has effectively eliminated matrix multiplication as a significant bottleneck. However, nonlinear layers, especially Softmax and GeLU, now represent the primary sources of overhead in MPC~\cite{wang2022characterization,li2022mpcformer,dong2023puma}. GeLU requires a high-order Taylor expansion involving many multiplications, while Softmax demands iterative squaring for exponentials and comparisons for numerical stability. Hence, we can see from Table~\ref{tab:overview} that most studies manage to improve the performance of Softmax and GeLU.
Apart from standard semi-honest assumption~\cite{li2022mpcformer,wang2022characterization,gupta2023sigma,luo2024secformer,chen2024securetlm,santos2024curl,dong2023puma}, PrivFormer~\cite{akimoto2023privformer} and PPTIF~\cite{liu2024pptif} can work under the honest-majority setup, where two out of three parties are honest adversaries, and the last one can be semi-honest or even malicious. Nonetheless, even though 2PC-Dealer and 3PC setups could improve secure inference performance efficiently, the assumption of a trusted dealer or non-colluding server parties is sometimes considered unrealistic in practice~\cite{hou2023ciphergpt}.

\begin{table}[t]
\begin{threeparttable}
  \caption{An overview of secure transformer inference studies with employed cryptographies.}
  \label{tab:overview}
  {\renewcommand{\arraystretch}{1.2}
  \small
  \begin{tabular}{l|l|l|c|c|c|c|>{\centering\arraybackslash}m{3cm}}
    \toprule 
    \multirow{2}{*}{Setup} & \multirow{2}{*}{Study} & \multirow{2}{*}{Tools} &\multicolumn{4}{c|}{Improved Components} & \multirow{2}{*}{Experiments on} \\\cline{4-7}
    & &  &MatMul &GeLU &Softmax &LayerNorm & \\\hline
    \multirow{22}{*}{2PC} & THE-X~\cite{chen2022x}  & HE  & $\circ$ & $\bullet$ & $\bullet$ & $\bullet$ &BERT-Tiny \\
    &Iron~\cite{hao2022iron}  & ASS+HE  & $\bullet$ &$\bullet$ & $\bullet$ &$\circ$ &BERT-Base \\
    &CipherGPT~\cite{hou2023ciphergpt}  &ASS+HE  &$\bullet$ &$\bullet$ &$\circ$ &$\circ$ &GPT2-Base\\
    &\multirow{2}{*}{Bumblebee~\cite{lu2023bumblebee}} & \multirow{2}{*}{ASS+HE}  & \multirow{2}{*}{$\bullet$} & \multirow{2}{*}{$\bullet$} & \multirow{2}{*}{$\bullet$} & \multirow{2}{*}{$\circ$} & BERT-Base/Large, GPT2- \\
    & & & & & & & Base, LLaMA-7B, ViT-Base \\
    &East~\cite{ding2023east} &ASS+HE  & $\bullet$ & $\bullet$ & $\bullet$ & $\bullet$ & BERT-Base\\
    &Zimerman~\emph{et~al.}~\cite{zimerman2023converting} &HE  &$\bullet$ &$\bullet$ &$\bullet$ &$\bullet$ & ViT, Swin Transformer\\
    &BOLT~\cite{pang2024bolt}  &ASS+HE  &$\bullet$ &$\bullet$ &$\bullet$ & $\circ$ &BERT-Base\\
    &NEXUS~\cite{zhang2024secure}  &HE  &$\bullet$ &$\bullet$ &$\bullet$ &$\bullet$ &BERT-Base, GPT-2\\
    &CipherFormer~\cite{wang2024cipherformer}   &HE+GC  &$\bullet$ &$\bullet$ &$\bullet$ &$\bullet$ & Transformer\\
   & MPCViT~\cite{zeng2023mpcvit}  &ASS  &$\bullet$ &$\bullet$ &$\bullet$ &$\circ$ &CCT, CVT \\
    &SAL-ViT~\cite{zhang2023sal}  &ASS  &$\bullet$ &$\circ$ &$\bullet$ &$\circ$ & CCT \\
    &\multirow{2}{*}{Primer~\cite{zheng2023primer}} &\multirow{2}{*}{HE+GC}  &\multirow{2}{*}{$\bullet$}  &\multirow{2}{*}{$\circ$}  &\multirow{2}{*}{$\bullet$}  & \multirow{2}{*}{$\circ$} &BERT-Tiny/Small/Base \\
    & & & & & & &/Medium/Large\\
    & PrivCirNet~\cite{xu2024privcirnet} &MPC+HE  &$\bullet$ &$\circ$ &$\circ$ &$\circ$ &ViT \\
    & \multirow{2}{*}{Liu~\emph{et~al.}~\cite{liu2023llms}} &\multirow{2}{*}{MPC+HE} &\multirow{2}{*}{$\bullet$} &\multirow{2}{*}{$\bullet$} &\multirow{2}{*}{$\bullet$} &\multirow{2}{*}{$\bullet$} &BERT-Tiny/Medium, \\
    & & & & & & &Roberta-Base\\
    & RNA-ViT~\cite{chen2023rna} &MPC  &$\bullet$ &$\bullet$ &$\bullet$ &$\circ$ &CCT \\
    & Powerformer~\cite{park2024powerformer} &HE  &$\bullet$ &$\bullet$ &$\bullet$ &$\bullet$ &BERT-Tiny \\
    & Rovida~\emph{et~al.}~\cite{rovida2024transformer} &HE  &$\bullet$ &$\bullet$ &$\bullet$ &$\bullet$ &BERT-Tiny \\
    & Zimerman~\emph{et~al.}~\cite{zimerman2024power} &HE  &$\bullet$ &$\bullet$ &$\bullet$ &$\circ$ &Roberta, GPT, ViT \\
    & SecureGPT~\cite{zeng2024securegpt} &MPC  &$\bullet$ &$\bullet$ &$\bullet$ &$\circ$ &GPT \\
    & SecBERT~\cite{huang2024secbert} &MPC  &$\bullet$ &$\bullet$ &$\bullet$ &$\circ$ &BERT \\
    & THOR~\cite{moon2024thor} &HE  &$\bullet$ &$\bullet$ &$\bullet$ &$\circ$ &BERT-Base \\
    \hline
    \multirow{6}{*}{2PC-Dealer}
    &MPCFormer~\cite{li2022mpcformer}  & MPC  &$\circ$ &$\bullet$ &$\bullet$ &$\circ$ &BERT-Base \\
    &Wang~\emph{et~al.}~\cite{wang2022characterization}  & ASS  & $\circ$ &$\circ$ & $\bullet$ & $\circ$ &XLM, ViT  \\
    &\multirow{3}{*}{SIGMA~\cite{gupta2023sigma}}  &\multirow{3}{*}{FSS}  & \multirow{3}{*}{$\circ$} &\multirow{3}{*}{$\bullet$} & \multirow{3}{*}{$\bullet$} & \multirow{3}{*}{$\bullet$} &BERT-Tiny/Base/Large, \\
    & & & & & & & GPT2/GPT-Neo,\\
    & & & & & & & Llama2-7B/13B\\
    &SecFormer~\cite{luo2024secformer}  &ASS  & $\circ$ &$\bullet$ & $\bullet$ & $\bullet$ &BERT-Base/Large\\
    &SecureTLM~\cite{chen2024securetlm}  &HSS  &$\bullet$ &$\bullet$ & $\bullet$ & $\bullet$ &-\\
    &\multirow{2}{*}{Curl~\cite{santos2024curl}}  &\multirow{2}{*}{LUT}  & \multirow{2}{*}{$\circ$} &\multirow{2}{*}{$\bullet$} & \multirow{2}{*}{$\bullet$} & \multirow{2}{*}{$\circ$} &BERT-Tiny/Base/Large\\
    & & & & & & & GPT2/GPT-Neo\\\hline
    \multirow{5}{*}{3PC}&\multirow{3}{*}{PUMA~\cite{dong2023puma}}  &\multirow{3}{*}{RSS} &\multirow{3}{*}{$\circ$} &\multirow{3}{*}{$\bullet$} &\multirow{3}{*}{$\bullet$} &\multirow{3}{*}{$\circ$} & BERT-Base/Large, GPT2-\\
    & & &  & & & & Base/Medium/Large, \\
    & & &  & & & & Roberta-Base, LLaMA-7B \\
    &PrivFormer~\cite{akimoto2023privformer}  &RSS  &$\circ$ &$\circ$ &$\bullet$ &$\circ$ & - \\
    &PPTIF~\cite{liu2024pptif}  &RSS &$\bullet$ &$\circ$ &$\bullet$ &$\circ$ &-\\
    
  \bottomrule
\end{tabular}
}
\end{threeparttable}
\end{table}

\subsection{Strengths and Weaknesses.}\label{subsec:pros}
MPC and HE both enable secure computation, and there is no clear better technique. Instead, the choice has to be made based on multiple factors depending on the applications. Here, we provide a critical review of their strengths and weaknesses in terms of communication cost, computational overhead and privacy protection.

\begin{table}[ht]
\begin{threeparttable}
  \caption{Strengths and weaknesses to current secure transformer inference studies.}
  \label{tab:pros}
  \begin{tabular}{p{2.2cm}|*{8}{>{\centering\arraybackslash}p{0.45cm}|}p{5cm}}
    \toprule
    \textbf{Study} & 
    \rotatebox{90}{\textbf{Overhead}} & \rotatebox{90}{\textbf{Client Computation }} & \rotatebox{90}{\textbf{Model Performance}} & 
    \rotatebox{90}{\textbf{Hardware Acceleration}} &
    \rotatebox{90}{\textbf{Client Data Privacy}} & 
    \rotatebox{90}{\textbf{Model Privacy}} & 
    \rotatebox{90}{\textbf{Scalability}} &
    \rotatebox{90}{\textbf{Code Availability}} &
    \textbf{Comments}\\ 
    \hline
    THE-X~\cite{chen2022x} &- & &- & &+ &- &- &- &Lack of strict privacy protection for intermediate inference results \\\hline
    MPCFormer~\cite{li2022mpcformer} &+ &- & & &+ &+ &- &+ & Require a trusted dealer and additional distillation training\\\hline
    Wang~\emph{et~al.}~\cite{wang2022characterization} &+ &- &? & &+ &+ & &- &Require a trusted dealer\\\hline
    Iron~\cite{hao2022iron} &- &- &+ & &+ &+ & &- & \\\hline
    CipherGPT~\cite{hou2023ciphergpt} & &- &+ & &+ &+ & &- &\\\hline
    Bumblebee~\cite{lu2023bumblebee} & &- &+ & &+ &+ &+ &- & \\\hline
    SIGMA~\cite{gupta2023sigma} &+ &- &+ &+ &+ &+ &+ &- &Require a trusted dealer \\\hline
    PUMA~\cite{dong2023puma} &+ &- &+ & &+ &+ & &+ &Operate under 3PC setup \\\hline
    PrivFormer~\cite{akimoto2023privformer} &- &+ &? & &+ & & &- &Operate under 3PC setup  \\\hline
    PPTIF~\cite{liu2024pptif} &- &+ &? & &+ &+ & &- &Operate under 3PC setup  \\\hline
    BOLT~\cite{pang2024bolt} & &- &+ &  &+ &+ & &- &\\\hline
    NEXUS~\cite{zhang2024secure} &- &+ &+ & &+ &+ & &- &Non-interactive protocol for inference \\
  \bottomrule
\end{tabular}
\end{threeparttable}
\end{table}

\textbf{Client Computation.} 
Some studies require the client to frequently participate in the computation during the inference process, and hence they tend to assume that clients also have relatively high computing resources. However, involving clients in computation could pose significant consumption challenges to low-power devices in practical, e.g., mobile devices. THE-X~\cite{chen2022x}

\textbf{Model Performance.}
THE-X~\cite{chen2022x} and MPCFormer~\cite{li2022mpcformer} simply replace the nonlinear layers with cryptography-friendly approximations for efficiency and, therefore, suffer a significant model performance degradation. To cure the accuracy drop, MPCFormer~\cite{li2022mpcformer} applies a knowledge distillation (KD)~\cite{hinton2006fast} method to train the approximated model, using teacher guidance from the original transformer.
However, the additional distillation period also brings extra computation overheads to the framework.

Later studies~\cite{hao2022iron,hou2023ciphergpt,lu2023bumblebee,gupta2023sigma,dong2023puma,pang2024bolt,zhang2024secure} have paid more attention to model inference performance. For the nonlinear layers, they often utilize look-up tables or high-order polynomial approximations to maintain accuracy while ensuring cryptographic compatibility. Hence, they can obtain satisfying performance on GLUE benchmarks~\cite{wang2018glue}, comparable to the original transformer. Stuides~\cite{wang2022characterization,akimoto2023privformer,liu2024pptif} focus more on the general inference framework of transformer encoder (decoder) and does not report performance results on specific intelligence tasks.

\textbf{Code Availability.} We provide links to the implementations of those open-source PTI research in Table~\ref{tab:codes}.

\begin{table}[t]
\begin{threeparttable}
  \caption{Links to available implementations.}
  \label{tab:codes}
  {\renewcommand{\arraystretch}{1.2}
  \small
  \begin{tabular}{l|c|c|c|c|c|c|p{3cm}}
    \toprule 
     Study & Year & Link &Comment \\\hline
     BumbleBee~\cite{lu2023bumblebee} & 2023 & \url{https://github.com/AntCPLab/OpenBumbleBee} & - \\
     BOLT~\cite{knott2021crypten} & 2024 & \url{https://github.com/Clive2312/BOLT} & - \\
     NEXUS~\cite{zhang2024secure} & 2024 & \url{https://github.com/zju-abclab/NEXUS} &- \\
     Zama-AI & - & \url{https://github.com/zama-ai/concrete-ml/tree/release/1.1.x/use_case_examples/llm} &- \\
  \bottomrule
\end{tabular}
}
\end{threeparttable}
\end{table}

\begin{table}[t]
\begin{threeparttable}
  \caption{An overview of PTI studies with employed MPC backbones.}
  \label{tab:overview}
  {\renewcommand{\arraystretch}{1.2}
  \small
  \begin{tabular}{l|c|c|c|c|c|c|p{3cm}}
    \toprule 
    \#Parties& Study & SS Scheme & MPC Engine &HE Method & HE Library  \\\hline
     \multirow{20}{*}{2PC} 
     &THE-X~\cite{chen2022x} &- &- &- &SEAL  \\
     &Iron~\cite{hao2022iron} &2-out-of-2 ASS &EzPC &BFV &SEAL  \\
    & CipherGPT~\cite{hou2023ciphergpt} &2-out-of-2 ASS &-  
    &BFV &SEAL\\
    & Bumblebee~\cite{lu2023bumblebee} &2-out-of-2 ASS &-  
    &BFV &SEAL\\
    & East~\cite{ding2023east} &2-out-of-2 ASS &EzPC  
    &BFV &SEAL\\
    & BOLT~\cite{pang2024bolt} &2-out-of-2 ASS &EzPC  
    &BFV &SEAL\\
    & NEXUS~\cite{zhang2024secure} &- &-  
    &RNS-CKKS &SEAL\\
    & MPCViT~\cite{zeng2023mpcvit} &2-out-of-2 ASS &SecretFlow-SPU &- &-\\
    & SAL-ViT~\cite{zhang2023sal} &2-out-of-2 ASS &CrypTen &- &-\\
    & SecFormer~\cite{luo2024secformer} &2-out-of-2 ASS &CrypTen &- &-\\
    & Primer~\cite{zheng2023primer} &2-out-of-2 ASS &- &- &-\\
    & PrivCirNet~\cite{xu2024privcirnet} &- &- &BFV &SEAL\\
    &RNA-ViT~\cite{chen2023rna} &2-out-of-2 ASS &CrypTen &- &-\\
    &Liu~\emph{et~al.}~\cite{liu2023llms} &2-out-of-2 ASS &OpenCheetah, SCI &BFV &-\\
    &PowerFormer~\cite{park2024powerformer} &- &- &RNS-CKKS &Lattigo\\
    &Rovida~\emph{et~al.}~\cite{rovida2024transformer} &- &- &RNS-CKKS &OpenFHE\\
    &Zimerman~\emph{et~al.}~\cite{zimerman2024power} &- &- &CKKS &HEaaN\\
    & SecureGPT~\cite{zeng2024securegpt} &- &emp-toolkit &- &- \\
    & SecBERT~\cite{huang2024secbert} &2-out-of-2 ASS &ABY &- &-\\
    & THOR~\cite{moon2024thor} &- &- &RNS-CKKS &Liberate.FHE\\
    \hline
    \multirow{6}{*}{2PC-Dealer} &MPCFormer~\cite{li2022mpcformer} &2-out-of-2 ASS &Crypten &- &-  \\
    &Wang~\emph{et~al.}~\cite{wang2022characterization}  & 2-out-of-2 ASS &Crypten &- &-\\
    &SIGMA~\cite{gupta2023sigma} & FSS &Orca &- &-\\
    &SecFormer~\cite{luo2024secformer} & 2-out-of-2 ASS &Crypten &- &-\\
    &SecureTLM~\cite{chen2024securetlm} & HSS &EzPC &- &-\\
    &Curl~\cite{gupta2023sigma} & LUT &- &- &-\\
    \hline
    \multirow{3}{*}{3PC} &PUMA~\cite{dong2023puma} &2-out-of-3 RSS &SecretFlow-SPU &- &-  \\
    & PrivFormer~\cite{akimoto2023privformer} &2-out-of-3 RSS &Falcon &- &-\\
    & PPTIF~\cite{liu2024pptif} &2-out-of-3 RSS &MP-SPDZ &- &-\\
  \bottomrule
\end{tabular}
}
\end{threeparttable}
\end{table}

\begin{table}[t]
\begin{threeparttable}
  \caption{Links to available backbones.}
  {\renewcommand{\arraystretch}{1.2}
  \small
  \begin{tabular}{l|c|c|c|c|c|c|p{3cm}}
    \toprule 
     Study & Year & Link &Comment \\\hline
     EzPC~\cite{chandran2019ezpc} & 2019 & \url{https://github.com/mpc-msri/EzPC} & - \\
     MP-SPDZ~\cite{keller2020mp} & 2020 & \url{https://github.com/data61/MP-SPDZ} & - \\
     Falcon~\cite{wagh2021falcon} & 2021 & \url{https://github.com/snwagh/falcon-public} & 3PC \\
     CrypTen~\cite{knott2021crypten} & 2021 & \url{https://github.com/facebookresearch/CrypTen} & - \\
     SecretFlow-SPU~\cite{ma2023secretflow} & 2023 & \url{https://github.com/secretflow/spu} &- \\
  \bottomrule
\end{tabular}
}
\end{threeparttable}
\end{table}

\section{Linear Layers in Transformers}\label{sec:linear}

Linear layers in transformers mainly involve large matrix multiplications (MatMul) in attention and feed-forward layers, and this section discusses the state-of-the-art literature on secure MatMul in PTI. 
Specifically, Section~\ref{subsec:linear_comp_transformer} first provides a breakdown of linear layers in a standard Transformer encoder. Then, Section~\ref{subsec:mat_mul} generalizes the two kinds of main-stream secure MatMul protocols in current PTI studies.

Based on the above analysis, we can see that high-dimensional \textbf{matrix multiplication (MatMul)} is the primary linear computation involved in transformers. Compared to the matrix-vector multiplications used in traditional networks (e.g., CNNs), MatMuls require significantly more computational resources due to the increased dimensionality and complexity~\cite{hao2022iron,pang2024bolt}. This is particularly evident in the attention mechanism: (\ref{equa:Q_h})-(\ref{equa:Atten_weights}). 
Thereafter, how to design efficient and secure MatMul protocols is crucial for PTI studies.

\subsection{Large Matrix Multiplications}\label{subsec:mat_mul}
The inference of a transformer-based model can involve hundreds of large matrix multiplications. 
For ease of understanding, we first present an example of MatMul operation as follows: 
\begin{example}
    Given two matrices $\boldsymbol{A}\in\mathbb{R}^{m\times n}$ and $\boldsymbol{B}\in\mathbb{R}^{n\times k}$, the matrix multiplication output $\boldsymbol{C}$ is:
    \begin{align}
        \bm{C}=\bm{A} \times \bm{B},
    \end{align}
    where $\bm{C}\in\mathbb{R}^{m\times k}$, and its each element $\boldsymbol{C}_{i,j}$ can be regarded as:
    \begin{align}
    \boldsymbol{C}_{i,j}= \sum_{k=1}^{N}\boldsymbol{A}_{i,k}\cdot \boldsymbol{B}_{k,j},\label{equa:matmul}
\end{align}
where $\boldsymbol{A}_{i,k}$ is an element from the $i$-th row of $\boldsymbol{A}$ and $\boldsymbol{B}_{k,j}$ is an element from the $j$-th column of $\boldsymbol{B}$.
\end{example}
According to (\ref{equa:matmul}), we can see that each element of the MatMul output $\boldsymbol{C}$ can be computed as the sum of products of corresponding elements from the rows of $\boldsymbol{A}$ and the columns of $\boldsymbol{B}$. This naive finding allows us to break the whole matrix multiplication into a series of addition and multiplication operations, and cryptographic primitives can be adapted to support these operations.

Particularly, to implement secure MatMul protocols, current PTI studies adopt the following main-stream ways:
\begin{itemize}
    \item \textbf{Secret Sharing (SS)-based Scheme}: 
    In the SS-based scheme, the matrices $\boldsymbol{A}$ and $\boldsymbol{B}$ are divided into shares and distributed among multiple parties. Each party holds a share of $\boldsymbol{A}$ and $\boldsymbol{B}$ but not the original matrices. The matrix multiplication is performed on the shares, and the results are combined to obtain the final result. 

    \textit{Strengths:}
    \begin{itemize}
        \item Additions can be easily evaluated locally ``for free'', dramatically saving communication resources.
    \end{itemize}
    \textit{Weaknesses:}
    \begin{itemize}
        \item 
        Multiplications need extra communication between parties. 
    \end{itemize}
    \item \textbf{Homomorphic Encryption (HE)-based Scheme}: 
    Here, one matrix ($\bm{A}$ or $\bm{B}$) is encrypted using HE that 
    supports arithmetic operations on ciphertexts. 
    The two matrices are then multiplied, and the MatMul output is decrypted to obtain the final product.
    
    \textit{Strengths:}
    \begin{itemize}
        \item HE-based methods naturally support mixed computations, which is communication-friendly.
    \end{itemize}
    \textit{Weaknesses:}
    \begin{itemize}
        \item Operations involving ciphertexts require more computation and are applied component-wisely by default, resulting in a considerable computational overhead. 
    \end{itemize}
\end{itemize}

Next, we further detail state-of-the-art PTI studies on SS-based MatMul protocols and HE-based MatMul protocols in Section~\ref{subsec:SS-based MatMul} and Section~~\ref{subsec:HE-based MatMul}, respectively.

\subsection{Secrete Sharing-based MatMul protocols}\label{subsec:SS-based MatMul}
We first introduce secrete sharing-based MatMul protocols in current secure transformer inference studies. 
Specifically, secrete sharing schemes support adding secret-shared values locally (without interaction) by simply adding the corresponding secret shares, making additions relatively inexpensive.
Hence, the key point is efficiently implementing multiplications while preserving privacy.
Various studies have explored different techniques for secure multiplication tailored to the specific MPC settings and SS mechanisms they employ.
Table~\ref{tab: SS-based MatMul} presents their utilized MPC settings, SS schemes and techniques for multiplications, along with a critical review of their strength and weaknesses in overheads. 

Iron~\cite{hao2022iron} builds its MatMul protocol on the 2PC setting and 2-out-of-2 ASS scheme. In this setup, direct multiplication would cause both parties to interact in a way that could reveal information about their shares. Iron leverages the Brakerski-Fan-Vercauteren (BFV) HE scheme to encrypt one party's share before communication, allowing the multiplication to be performed on encrypted data (ciphertexts). To manage the computational complexity, Iron packs the plaintext matrices into polynomials before applying HE encryption. 
This packing strategy is based on the insight that, by arranging the polynomial coefficients in a specific manner, the resulting polynomial multiplication corresponds directly to the desired MatMul results~\cite{huang2022cheetah}.
Similar to Iron, BumbleBee~\cite{lu2023bumblebee} also combines ASS with HE and starts with the coefficient encoding approach. However, it further improves the packing scheme with a procedure named \textit{IntrLeave}, which could efficiently interleave multiple polynomials into one, reducing the size of ciphertexts for transmission. Consequently, it requires much less communication compared to Iron. CipherGPT~\cite{hou2023ciphergpt} accelerates MatMuls by leveraging a preprocessing phase with sVOLE~\cite{boyle2018compressing} correlations and batching multiple matrix multiplications together, reducing both computation and communication consumptions. 


MPCFormer~\cite{li2022mpcformer} and Wang~\emph{et~al.}~\cite{wang2022characterization} operate under the 2-out-of-2 ASS but introduces an additional trusted third-party dealer. The role of the dealer is to provide a secret shared triple, $c=ab$, to two parties and facilitate the multiplication operation, and such a joint computation technique is called Beaver triple~\cite{beaver1992efficient}.
The introduction of Beaver triple requires one extra round of communication for multiplication compared to direct multiplication without MPC. However, neither MPCFormer nor Wang~\emph{et~al.}~\cite{wang2022characterization} uses other additional techniques to accelerate matrix multiplication, and they both claim that the nonlinear layers in Transformers are the major sources of bottlenecks. In particular, matrix multiplications account for only 12.7\% of the overall BERT-Base model runtime under MPC, and the communication overhead is also much lower than that of the non-linear layer.
Privformer~\cite{akimoto2023privformer} and PUMA~\cite{dong2023puma} build their MatMul protocols upon the setup of 3PC and 2-out-of-3 RSS scheme. All three parties are directly involved in the MPC computations. 
For multiplication, each of the three parties computes part of the result locally, then exchanges noisy results via the \textit{re-sharing} method in Araki~\emph{et~al.}~\cite{araki2016high}, and finally combines them into a new shared product value. 

Generally, SS-based MatMul approaches above mainly rely on either 2PC+HE~\cite{hao2022iron,lu2023bumblebee,hou2023ciphergpt} or introduce an additional helper party to participate in computations~\cite{li2022mpcformer,wang2022characterization,dong2023puma,akimoto2023privformer}.
The former is more convenient for practical application in secure inference, as the server and the client can act as two computational parties, eliminating the need for additional assumptions about parties. However, 2PC still poses a very large overhead due to using additional cryptography to support multiplication~\cite{rathee2024mpc}. In contrast, 2PC-Dealer and 3PC setups are effective in improving the efficiency of MatMuls, but they rely on a trusted dealer or honest-majority assumption. 
It is difficult to compare the two approaches in a strict sense because of their disparities in the security assumption and applicable scenarios.

\begin{table}
\begin{threeparttable}
  \caption{A Comparison of SS-based MatMul Protocols.}
  \label{tab: SS-based MatMul}
  \begin{tabular}{l|l|l|l|c|c|l}
    \toprule
    Study & MPC Setup &SS Scheme &${\textnormal{Tech. for Mul.}}^{1}$ & ${\textnormal{Comm.}}^{2}$ &${\textnormal{Comp.}}^{3}$ & Comments\\
    \midrule
    
    Iron~\cite{hao2022iron} &2PC  & 2-out-of-2 ASS & HE  &- & & \\
    BumbleBee~\cite{lu2023bumblebee} & 2PC  & 2-out-of-2 ASS & HE  & & & \\
    CipherGPT~\cite{hou2023ciphergpt} & 2PC & 2-out-of-2 ASS & HE  & &+ & \\
    MPCFormer~\cite{li2022mpcformer} &2PC-Dealer  & 2-out-of-2 ASS & Beaver triple  & & &Needs a trusted dealer \\
    Wang~\emph{et~al.}~\cite{wang2022characterization} &2PC-Dealer  & 2-out-of-2 ASS & Beaver triple  & & &Needs a trusted dealer \\
    PUMA~\cite{dong2023puma} &3PC  & 2-out-of-3 RSS & re-sharing  &+ &+ & Honest-majority \\ 
    Privformer~\cite{akimoto2023privformer} & 3PC  & 2-out-of-3 RSS & re-sharing  &+ &+ &Honest-majority \
    \\
  \bottomrule
\end{tabular}
\begin{tablenotes}
	\small {
     \item ${}^{1}$ Techniques for Multiplications. \hspace{0.5em}
     ${}^{2}$ Communication.
     \hspace{0.5em}
     ${}^{3}$ Computation.
     \item  Note that + indicates a strength and - to a weakness. A blank field means a solution is neither strong nor weak in this area.
            }
    \end{tablenotes}
\end{threeparttable}
\end{table}
\subsection{Homomorphic Encryption-based MatMul protocols}\label{subsec:HE-based MatMul}
The use of HE for matrix multiplications in transformers is still relatively unexplored.
There is little work on PTI using HE-based protocols for matrix multiplications.
While matrix multiplication is inherently compatible with HE, as it involves only additions and multiplications, a straightforward implementation can be highly inefficient. 
Specifically, HE-based MatMul protocols in transformers still face several key challenges: 
\begin{itemize}
    \item \textbf{Transmission Overheads:} Ciphertext matrices are considerably larger than plaintext matrices, leading to substantial transmission costs.
    \item \textbf{Computational Expense:} A vast number of multiplications are required, and these multiplications are computationally expensive in the context of HE.
    \item \textbf{Limited Multiplicative Depth:} The multiplicative depth in HE is constrained by the accumulation of noise, often necessitating expensive bootstrapping operations to refresh the depth and enable further computations.
\end{itemize}
These deficiencies motivate the need for more efficient HE-based MatMul protocols, and we report the comparison of HE-based MatMul protocols in Table~\ref{tab:HE-based MatMul}.

An early study, THE-X~\cite{chen2022x}, first proposes to utilize an FHE scheme, CKKS, to implement secure transformer inference. It naively converts all matrix multiplications into the element-wise style and employs no other acceleration techniques. Hence, it only evaluates a small transformer model BERT-Tiny and does not report running time performance.

BOLT~\cite{pang2024bolt} supports MatMuls in transformers using a leveled FHE scheme, BFV. It employs a column-wise packing for the ciphertext matrix and adopts Gazelle's diagonal packing~\cite{juvekar2018gazelle} for the plaintext matrix. 
Ciphertext packing is to organize and pack multiple plaintext data into a single ciphertext, which allows for parallel operations using SIMD techniques. 
BOLT also utilizes the outer product representation and partial rotation techniques to speed up the ciphertext-ciphertext MatMuls (e.g., $\boldsymbol{Q}\cdot\boldsymbol{K}^{T}$ in the attention layer). 
BOLT operates under an MPC-HE hybrid framework, which allows it to automatically reset the HE noise without bootstrapping operations.

The latest study,
NEXUS~\cite{zhang2024secure}, operates with RNS-CKKS and incorporates SIMD ciphertext compression. This technique compresses multiple ciphertexts into a single one, thereby reducing the volume of data transferred. Decompression is achieved with a fixed number of operations, adding no additional computational overhead. Additionally, NEXUS employs SIMD slots folding, accelerating ciphertext-ciphertext MatMuls. This is accomplished by efficiently aggregating element-wise products within a ciphertext using significantly fewer rotations, thus enhancing processing speed.

\begin{table}[h]
\begin{threeparttable}
  \caption{A Comparison of HE-based MatMul Protocols.}
  \label{tab:HE-based MatMul}
  \begin{tabular}{l|l|c|c|c|c|c|l}
    \toprule
    Study &HE Scheme &Packing & SIMD & Bootstrapping & $\textnormal{Comm.}^{1}$ & $\textnormal{Comp.}^{2}$ &Comments \\
    \midrule
    THE-X~\cite{chen2022x} & CKKS & $\times$ & $\times$ & ? &- &- &Tiny evaluation model  \\
    BOLT~\cite{pang2024bolt} & BFV & \checkmark & \checkmark & $\times$ &+ &+ &  \\
    NEXUS~\cite{zhang2024secure} &  RNS-CKKS & \checkmark & \checkmark &\checkmark &+ & & \\
  \bottomrule
\end{tabular}
\begin{tablenotes}
	\small {
     \item 
     ${}^{1}$ Communication.
     \hspace{0.5em}
     ${}^{2}$ Computation.
     \item  Note that + indicates a strength and - to a weakness. A blank field means a solution is neither strong nor weak in this area. And ? means the information is missing in the study.
            }
    \end{tablenotes}
\end{threeparttable}
\end{table}

Compared to secrete-sharing based MatMuls, 
HE often allows for more savings on communication bandwidth and roundtrips required for computation. However, computations on ciphertexts are computationally expensive, and large MatMuls in transformers could further increase multiplicative depth, complicating ciphertext-ciphertext matrix multiplication. 
Current studies~\cite{pang2024bolt,zhang2024secure} mainly leverage ciphertext packing schemes and SIMD techniques to address the above challenges. 
This approach significantly reduces the number of ciphertexts and speeds up computations by enabling efficient parallel processes.









\section{Non-linear Layers}\label{sec:non_linear_layers}
Securing the non-linear functions in Transformers poses another challenge due to their complexity in cryptographic primitives. The bulk of the total run time is from non-linear layers for both MPC and HE studies~\cite{hao2022iron,li2022mpcformer,pang2024bolt}. Non-linear functions in Transformers include Softmax, GeLU and LayerNorm. We detail each of them in the following subsections.

\subsection{Softmax}\label{subsec:softmax}
In the attention mechanism, one key challenge is the implementation of the nonlinear Softmax operation.
For the sake of numerical stability~\cite{goodfellow2016deep}, when given an input vector $\boldsymbol{x} \in \mathbb{R}^d$, the Softmax function can be computed as:
\begin{align}
    \mathrm{Softmax}(\boldsymbol{x})=\left[\frac{\exp{(x_i-\hat{x})}}{\sum_{j=1}^{d}\exp{(x_j-\hat{x})}}\biggm|_{i\in[d]}\right],\label{equa:softmax}
\end{align}
where $\hat{x}=\max_{i\in [d]}x_i$. It is worth noting that all inputs to the exponential function in (\ref{equa:softmax}) are non-positive now.
The main challenge is to efficiently calculate the underlying exponential function and division computation. Next, we discuss recent research advancements in optimizing Softmax and provide a detailed comparison in Table~\ref{tab:softmax}.

\begin{tcolorbox}[tech-tip]
    \textbf{Tech Tips:} In essence, the $\mathrm{Softmax}$ function could be regarded to perform a re-weight normalization of the obtained attention map, and its general form is formulated as follows:
    \begin{align}
        \mathrm{Softmax}(x_i)=\frac{F(x_i)}{\sum_{j=1}^d F(x_j)+\epsilon},
    \end{align}
    where function $F(\cdot)$ widely adopts the exponential function $\exp(\cdot)$ and $\epsilon$ is a small positive number to avoid a zero denominator. Therefore, existing studies mainly optimize its computation in the following two methods:
    \begin{itemize}
        \item Employ more crypto-friendly functions to server as $F(x)$ in Softmax, such as the $\mathrm{ReLU}(x)$ and  quadratic function $(x+c)^2$. Those functions often consist of more basic arithmetic operations (e.g., multiplications and comparisons), the overhead to compute $\mathrm{Softmax}(x_i)$, thus dramatically decreasing overheads. It is worth noting that this approach comes at the cost of a significant drop in accuracy.
        \item Directly utilize different polynomial approximations (e.g., the Taylor series) of $\mathrm{exp}(x)$ to serve as $F(x)$. This method
        could help to maintain the accuracy. However, one should be careful about the choice of polynomial approximation order; too high a polynomial order can bring better accuracy while reducing computational efficiency obviously.
    \end{itemize}
\end{tcolorbox}

Next, we classify the selected PTI studies based on the Tech Tips above and discuss how they cope with the Softmax function in more detailedly.

\textbf{Crypto-friendly Functions.} 
Studies~\cite{li2022mpcformer,zeng2023mpcvit,zhang2023sal,chen2023rna,luo2024secformer} that favour the first solution often use aggressive crypto-friendly functions and, therefore, require additional training to minimize the loss of accuracy. Specifically, we present their methods as follows:
\begin{align}
    \mathrm{Softmax}(x_i) \approx \frac{F(x_i)}{\sum_{j=1}^d F(x_j)+\epsilon},
    ~\textnormal{for}~F(x)=
    \begin{cases}
        (x+c)^2,&\textnormal{in MPCFormer~\cite{li2022mpcformer}, SecFormer~\cite{luo2024secformer}}\\
        \mathrm{ReLU}(x), &\textnormal{in MPCViT~\cite{zeng2023mpcvit}}\\
        (a\cdot x +c)^2,&\textnormal{in SAL-ViT~\cite{zhang2023sal}}\\
        (x +c)^4,&\textnormal{in RNA-ViT~\cite{chen2023rna}}\\
    \end{cases} \label{equa:appox_softmax1}
\end{align}
It is obvious that (\ref{equa:appox_softmax1}) and the original Softmax (\ref{equa:softmax}) differ a lot by numerical values. Hence, the above studies all utilize the Knowledge Distillation (KD)~\cite{hinton2006fast} method to bridge the performance gap.

\textbf{Polynomial Approximations.} 
Studies~\cite{pang2024bolt,zhang2024secure,hou2023ciphergpt,lu2023bumblebee,dong2023puma} employ polynomial approximation to replace the original $\exp(x)$ in Softmax. The computation of polynomials can be decomposed into basic addition and multiplication operations, which are easily supported by cryptographic techniques.
BOLT~\cite{pang2024bolt} decomposes the input $x_i-\hat{x}$ into a non-negative integer $z$ and a float number $p \in (-\ln2,0]$, where $x_i-\hat{x} = (-\ln 2)\cdot z + p$. Then, the exponentiation is approximated as $\exp(\widetilde{x}_i)=\exp (p) \gg z$, where $\gg$ denotes the right shift operator. Since the range of $p$ is relatively small, a 2-degree polynomial is enough for the approximation:
\begin{align}
    \exp(p) \approx 0.3585(p+1.353)^2+0.344.
\end{align}
For the $\frac{1}{\sum_{j \in[d]}\exp{(x_j-\hat{x})}}$ part, i.e., the reciprocal of the summed exponential function's outputs, BOLT uses the reciprocal protocol from SIRNN~\cite{rathee2021sirnn}. Besides, 
NEXUS~\cite{zhang2024secure}, CipherGPT~\cite{hou2023ciphergpt}, PUMA~\cite{dong2023puma} and BumbleBee~\cite{lu2023bumblebee} all approximate the exponential function in Softmax using the Taylor series. 
\begin{align}
    \exp(x) \approx
    \begin{cases}
        0, &\textnormal{if}~x \leq a\\
        (1+\frac{x}{2^r})^{2^r}, &\textnormal{if}~a\leq x \leq 0
    \end{cases},
    ~\textnormal{for}
    \begin{cases}
        a=-\inf,r=6,&\textnormal{in NEXUS~\cite{zhang2024secure}}\\
        a=-13,r=6,&\textnormal{in BumbleBee~\cite{lu2023bumblebee}}\\
        a=-16,&\textnormal{in CipherGPT~\cite{hou2023ciphergpt}}\\
        a=-14,r=5,&\textnormal{in PUMA~\cite{dong2023puma}}
    \end{cases}
\end{align}
Additionally, NEXUS uses its proposed $\mathrm{QuickSum}$ and the Goldschmidt division algorithm to compute the reciprocal of the sum of exponential functions. Rovida~\emph{et~al.}~\cite{rovida2024transformer} approximates Softmax using the Maclaurin series:
\begin{align}
    \mathrm{exp}(x)\approx\sum_{i=0}^6\frac{x^i}{i!},
\end{align}
and for the division operation, they use a Chebyshev polynomial approximation:
    \begin{align}
        \frac{1}{x} \approx \frac{c_0}{2}+\sum_{i=1}^{119}c_iT_i(x).
    \end{align}

\textbf{LUTs.} 
Some studies~\cite{hao2022iron,gupta2023sigma} use look-up tables (LUTs) for Softmax. A LUT is a pre-computed table stored in memory, containing input-output pairs. It allows for quick retrieval of function values by directly looking up results, eliminating the need for real-time calculation.
Iron~\cite{hao2022iron} uses a tree-reduction protocol to evaluate the max function and then invokes the LUT-based exponential and reciprocal protocols~\cite{rathee2021sirnn} to evaluate Softmax.
Sigma~\cite{gupta2023sigma} optimizes Softmax by introducing efficient methods for calculating the max, negative exponential, and inverse functions using reduced bit-width LUTs, significantly improving computational efficiency and reducing resource consumption. Generally speaking, LUTs can significantly reduce the computation time but typically incur a higher communication overhead, which can be validated in Table~\ref{tab:softmax} (e.g., Sigma~\cite{gupta2023sigma} to BumbleBee~\cite{lu2023bumblebee}, Iron~\cite{hao2022iron} to BOLT~\cite{pang2024bolt}).




\begin{table}[ht]
\begin{threeparttable}
  \caption{A Comparison of Softmax Approximations.}
  \label{tab:softmax}
  \begin{tabular}{c|c|c|c|c|c|c|c}
    \toprule
    Study &Input &Comm. (MB) & Comm. set. &Runtime (s) & Error & Comments\\
    \midrule
    MPCFormer~\cite{li2022mpcformer}  &- & -& -&4.9  &-&Require distillation\\\hline
    THE-X~\cite{chen2022x} &- & -& -& -& - &Require training\\\hline
    PUMA~\cite{dong2023puma} &- & -& -& -& -&No breakdown provided\\\hline
    BumbleBee~\cite{lu2023bumblebee}  & $\mathbb{R}^{128\times128}$ & 162.24 & $\textnormal{LAN}^{1}/\textnormal{WAN}^{2}$& $2.11/5.79$ &$ < 2^{-10}$ &-\\\hline
    CipherGPT~\cite{hou2023ciphergpt}  & $\mathbb{R}^{256\times256}$ & 277.59 & $\textnormal{LAN}^{3}$&  $14.865$ &- &-\\\hline
    Sigma~\cite{gupta2023sigma} &  $\mathbb{R}^{128\times128}$ & 266.24 & $\textnormal{LAN}^{4}$ & $0.44$ &- &-\\\hline
    Iron~\cite{hao2022iron} & $\mathbb{R}^{128\times128}$ &  3596.32 & $\textnormal{LAN}^{3}/\textnormal{WAN}^{5}$ & $60/1900$ &$3.2\times10^{-5}$ &-\\\hline
    BOLT~\cite{pang2024bolt} & $\mathbb{R}^{128\times128}$ &  1447.65 & $\textnormal{LAN}^{3}/\textnormal{WAN}^{5}$  & $16/775$ & $1.4\times10^{-6}$ &-\\\hline
    NEXUS~\cite{pang2024bolt} & $\mathbb{R}^{128\times128}$ &  0 & $\textnormal{LAN}^{3}/\textnormal{WAN}^{5}$  & $242/242$ &$3.1\times10^{-6}$ & Nearly comm. free \\
  \bottomrule
\end{tabular}
\begin{tablenotes}
	\small {
	 \item ${}^{1}$1Gbps, 0.5ms \hspace{0.5em}
            ${}^{2}$400Mbps,4ms \hspace{0.5em}
            ${}^{3}$3Gbps, 0.8ms \hspace{0.5em}
            ${}^{4}$9.4Gbps, 0.05ms
            \hspace{0.5em}
            ${}^{5}$100Mbps, 80ms 
            }
    \end{tablenotes}
\end{threeparttable}
\end{table}

\subsection{GeLU}\label{subsec:GeLU}
Rather than crypto-friendly ReLU, Transformers utilize GeLU activations. The GeLU (Gaussian Error Linear Unit) activation function is defined as follows:
\begin{align}
    \mathrm{GeLU}(x)=\frac{x}{2}\big(1+\mathrm{erf}(\frac{x}{\sqrt{2}})\big),
\end{align}
where $\mathrm{erf}(\cdot)$ denotes the Gaussian error function which is given by  $\mathrm{erf}(x)=\frac{2}{\sqrt{\pi}}\int_0^xe^{-t^2}dt$.
In many libraries, e.g., PyTorch, the following approximation is provided:
\begin{align}
    \mathrm{GeLU}(x) \approx \frac{x}{2}\Big(1+\mathrm{tanh}\big(\sqrt{\frac{2}{\pi}}(x+0.044715x^3)\big)\Big).
\end{align}
and where:
\begin{align}
    \mathrm{tanh}(x)=\frac{e^{2x}-1}{e^{2x}+1}.
\end{align}
Hence, the main challenge is how to efficiently handle the nonlinear $\mathrm{erf}(x)$ (i.e., $\mathrm{tanh}(x)$) part.

\begin{tcolorbox}[tech-tip]
    \textbf{Tech Tips:} The function $\mathrm{GeLU}(x)$ . Besides, polynomials are still available to approximate either $\mathrm{tanh}(x)$ (i.e., $\mathrm{erf}(x)$) or the entire $\mathrm{GeLU}(x)$
\end{tcolorbox}

Some studies~\cite{chen2022x,li2022mpcformer} approximate GeLU simply using crypto-friendly ReLU or ReLU approximation. 
For example, THE-X~\cite{chen2022x} and PowerFormer~\cite{park2024powerformer} directly replaces the GeLU layers with ReLU activation functions:
\begin{align}
    \mathrm{GeLU}(x) \approx \mathrm{ReLU}(x)= \max(0,x),
\end{align}
where the computation of function $\max(\cdot)$ is left to the user to complete after decryption.
Obviously, such an aggressive approximation is not accurate enough.
MPCFormer~\cite{li2022mpcformer} takes a quadratic function to approximate GeLU:
\begin{align}
    \mathrm{GeLU}(x) \approx0.125x^2+0.25x+0.5.\label{equa:GeLU_MPCFormer}
\end{align}
However, Equa.~(\ref{equa:GeLU_MPCFormer}) is designed for ReLU approximation in Chou~et~al.~\cite{chou2018faster} and thus also introduces a significant accuracy drop (MPCFormer mitigates this drop by distillation).

\begin{align}
    \mathrm{GeLU}(x) \approx
    \begin{cases}
        \max(0,x),&\textnormal{in THE-X~\cite{chen2022x}, PowerFormer~\cite{park2024powerformer}}\\
        0.125x^2+0.25x+0.5,&\textnormal{in MPCFormer~\cite{li2022mpcformer}}\\
        \max(0,x)+a*\min(0,x),&\textnormal{in RNA-ViT~\cite{chen2023rna}}
    \end{cases}
\end{align}

Some studies~\cite{hao2022iron,gupta2023sigma,hou2023ciphergpt} utilize LUT-based methods for GeLU.
Iron~\cite{hao2022iron} notices that the sign of $\mathrm{tanh}(x)$ is equal to that of $x$. Hence, it first generates a negative input $\overline{x}$ where $\abs{\overline{x}}=\abs{x}$, and then the following two cases hold: $\mathrm{tanh}(x)=\mathrm{tanh}(\overline{x})$ for $x\leq0$ and $\mathrm{tanh}(x)=-\mathrm{tanh}(\overline{x})$ for $x>0$.
Then, similar to dealing with Softmax, it calls the LUT-based protocols from SIRNN~\cite{rathee2021sirnn} to cope with the tanh function.
SIGMA~\cite{gupta2023sigma} finds the near-linearity and symmetry of GeLU, and it further employs a function $\delta(x)$ to measure the difference  between $\mathrm{GeLU}(x)$ and $\mathrm{ReLU}(x)$ for $x\in[-4,4]$:
\begin{align}
    \mathrm{GeLU}(x) \approx 
    \begin{cases}
        0,&\textnormal{if}~x<-4\\
        \mathrm{ReLU}(x)-\delta(x),&\textnormal{if}~-4\leq x \leq4\\
        x,&\textnormal{if}~x>4
    \end{cases}
\end{align}
where $\mathrm{ReLU}(x)$ is implemented by its proposed DReLU protocol, and $\sigma(x)$ is computed through the LUT.

\begin{table}
\begin{threeparttable}
  \caption{A Comparison of GeLU Approximations.}
  \label{tab:GeLu}
  \begin{tabular}{c|c|c|c|c|c|c}
    \toprule
    Study  & Input & Comm. (MB) & Comm. set. &Runtime (s) & Error & Comments\\
    \midrule
    THE-X~\cite{chen2022x} &-&-&-&-&-&Require  training\\\hline
    MPCFormer~\cite{li2022mpcformer} &-&-&-&-&-&Require  distillation\\\hline
    PUMA~\cite{dong2023puma} &- & -& -& -& -&No breakdown provided\\\hline
    BumbleBee~\cite{lu2023bumblebee}  & $\mathbb{R}^{128\times3072}$ & 162.24 & $\textnormal{LAN}^{1}/\textnormal{WAN}^{2}$& $2.11/5.79$ &$ < 2^{-10}$ &-\\\hline
    CipherGPT~\cite{hou2023ciphergpt}  & $\mathbb{R}^{256\times3072}$ & 575.91 & $\textnormal{LAN}^{3}$&  $20.657$ &- &-\\\hline
    SIGMA~\cite{gupta2023sigma} &$\mathbb{R}^{128\times3072}$ &163.84 &$\textnormal{LAN}^{4}$ &0.25 &-&\\\hline
    Iron~\cite{hao2022iron} &$\mathbb{R}^{128\times3072}$ & 7960.00 & $\textnormal{LAN}^{3}/\textnormal{WAN}^{5}$ & $126/4118$ &$5.8\times10^{-4}$& \\\hline
    BOLT~\cite{pang2024bolt} & $\mathbb{R}^{128\times3072}$ & 1471.67 & $\textnormal{LAN}^{3}/\textnormal{WAN}^{5}$ & $14/774$ &$9.8\times10^{-4}$ & \\\hline
    NEXUS~\cite{pang2024bolt} & $\mathbb{R}^{128\times3072}$ & 0 & $\textnormal{LAN}^{3}/\textnormal{WAN}^{5}$ & $233/233$ & $7.7\times10^{-4}$& Nearly comm. free\\
  \bottomrule
\end{tabular}
\begin{tablenotes}
	\small {
	 \item ${}^{1}$1Gbps, 0.5ms \hspace{0.5em}
            ${}^{2}$400Mbps,4ms \hspace{0.5em}
            ${}^{3}$3Gbps, 0.8ms \hspace{0.5em}
            ${}^{4}$9.4Gbps, 0.05ms
            \hspace{0.5em}
            ${}^{5}$100Mbps, 80ms 
            }
    \end{tablenotes}
\end{threeparttable}
\end{table}

Some studies~\cite{dong2023puma,zhang2024secure,pang2024bolt,lu2023bumblebee} employ polynomials to approximate GeLU efficiently.
PUMA~\cite{dong2023puma}, BumbleBee~\cite{lu2023bumblebee} and NEXUS~\cite{zhang2024secure} notice that $\mathrm{GeLU}$ function is almost linear on the two sides, i.e., $\mathrm{GeLU}(x) \approx 0$ for $x<-4$ and $\mathrm{GeLU}(x) \approx x$ for $x>3$. 
They also suggest an efficient low-degree polynomial for approximation within the short interval around $0$:
\begin{align}
    \mathrm{GeLU}(x)\approx
    \begin{cases}
        -c,&\textnormal{if}~x<a\\
         \sum_{i=0}^3a_ix^i, &\textnormal{if}~ a < x \leq b\\
        \sum_{i=0}^6b_ix^i, &\textnormal{if}~ b < x \leq 3\\
        x-c,&\textnormal{if}~x > 3\\
    \end{cases}
    ,~\textnormal{for}
    \begin{cases}
        a=-4,b=-1.95,c=0,&\textnormal{in PUMA~\cite{dong2023puma}, NEXUS~\cite{zhang2024secure}}\\
        a=-5,b=-2,c=10^{-5},&\textnormal{in BumbleBee~\cite{lu2023bumblebee}}
    \end{cases}
\end{align}
Similarly, BOLT~\cite{pang2024bolt} presents an approximation using a 4-degree polynomial:
\begin{align}
    \mathrm{GeLU}(x)\approx
    \begin{cases}
        x,&~\textnormal{if}~x > 2.7\\
        \sum_{i=0}^4a_i\abs{x}^i
           +0.5x,
        &~\textnormal{if}~\abs{x} \leq 2.7\\
        0,&~\textnormal{otherwise}\\
    \end{cases}
\end{align}

SecFormer~\cite{luo2024secformer} introduces the Fourier series to approximate the $\mathrm{erf}(x)$:
\begin{align}
    \mathrm{erf}(x)\approx \begin{cases}
        -1,&~\textnormal{if}~x < -1.7\\
        \boldsymbol{\beta} \odot \sin{(\boldsymbol{k} \odot \pi x/10)},
        &~\textnormal{if}~-1.7 \leq x \leq 1.7\\
        1,&~\textnormal{otherwise}\\
    \end{cases}
\end{align}


\subsection{Layer Normalization}\label{subsec:LayerNorm}
Layer normalization ensures that the inputs across different layers of the network have a consistent mean and variance, helping to stabilize deep neural networks. It is applied after each attention block and feed-forward layer.
For a given vector $\boldsymbol{x}\in \mathbb{R}^d$, the Layer normalization function is defined as follows:
\begin{align}
    \mathrm{LayerNorm}(\boldsymbol{x}) = \Big[\frac{\gamma(x_i-\mu)}{\sigma}+\beta \mid_{i=1,\ldots,d}\Big],
\end{align}
where $\mu=\sum_{i=1}^dx_i/d$ and $\sigma=\sqrt{\sum_{i=1}^d(x_i-\mu)^2}$ are mean and standard deviation, and $\gamma$ and $\beta$ are affine transform parameters. The calculation of $\mu$ only requires additions and multiplications by a constant. Thus, the main challenge lies in the required reciprocal square root operation. Next, we discuss the methods to cope with LayerNorm in different studies and report the performance in Tab.~\ref{tab:LayerNorm}.

Some studies~\cite{chen2022x,hao2022iron,zhang2024secure} have developed methods to mitigate the computation overhead of parameters $\mu$ and $\beta$ for efficiency.
For instance, THE-X~\cite{chen2022x} employs two sets of learnable parameters $\widetilde{\boldsymbol{\gamma}}:=[\widetilde{\gamma}_1,\ldots,\widetilde{\gamma}_d]$ and $\widetilde{\boldsymbol{\beta}}:=[\widetilde{\beta}_1,\ldots,\widetilde{\beta}_d]$ to avoid the calculations of original $\mu$ and $\beta$. Specifically, it simplifies $\mathrm{LayerNorm}$ as follows:
\begin{align}
    \mathrm{LayerNorm}(\boldsymbol{x}) \approx \boldsymbol{x} \odot \boldsymbol{\gamma} + \boldsymbol{\beta},\label{equa:layernorm_THEX}
\end{align}
where $\widetilde{\boldsymbol{\gamma}}$ handles both normalization and linear transformation, and $\odot$ denotes for the Hadamard product. A distillation will be applied in (\ref{equa:layernorm_THEX}) to learn from the original LN layers and bridge the performance gap.
Iron~\cite{hao2022iron} combines the weights of $\mathrm{LayerNorm}$ $(\frac{\gamma}{\sigma}$ and $\beta$) with the next linear layer's weights to save one matrix multiplication. This results in a simplified output from $\mathrm{LayerNorm}$ as follows:
\begin{align}
    \mathrm{LayerNorm}(\boldsymbol{x}) \approx \Big[x_i-\mu\mid_{i=1,\ldots,d}\Big].\label{equa:layernorm_iron}
\end{align}
However, BOLT~\cite{pang2024bolt} points out that Iron ignores the residual connection in transformers and using Equa.~(\ref{equa:layernorm_iron}) for LayerNorm would make the model’s
performance close to random guessing.
Besides, NEXUS~\cite{zhang2024secure} performs the following transformation to $\mathrm{LayerNorm}$ :
\begin{align}
    \mathrm{LayerNorm}(\boldsymbol{x}) = \left[\sqrt{d}\gamma \cdot \frac{z_i}{\sqrt{{\sum_{i=1}^dz_i^2}}}+\beta \biggm|_{i=1,\ldots,d}\right],
\end{align}
where $z_i=dx_i-\sum_{i=1}^dx_i$.
Then, NEXUS uses its proposed $\mathrm{QuickSum}$ function and the Newton iteration method in Qu~et al.~\cite{qu2023improvements}  to compute the summations and the inverse square root, respectively. Similar to NEXUS, PowerFormer~\cite{park2024powerformer} approximates $\frac{1}{\sqrt{x}}$ using an iterative algorithm based on Newton's method.

Some studies~\cite{gupta2023sigma,hou2023ciphergpt,akimoto2023privformer} employ customized protocols for the reciprocal square root operation in LayerNorm layers. 
CipherGPT~\cite{hou2023ciphergpt} directly introduces an LUT to evaluate the reciprocal square root operation. Similarly, SIGMA~\cite{gupta2023sigma} designs a custom 13-bit floating-point representation for efficiency to encode the input of LayerNorm and use it to index a LUT for the reciprocal square root. PrivFormer~\cite{akimoto2023privformer} utilizes the square-root inverse protocol in FALCON~\cite{li2020falcon} for LayerNorm and further improves its security under MPC settings.

\begin{table}
\begin{threeparttable}
  \caption{A Comparison of LayerNorm Approximations.}
  \label{tab:LayerNorm}
  \begin{tabular}{c|c|c|c|c|c|c}
    \toprule
    Study  & Input &Comm. (MB) & Comm. set. &Runtime (s) & Error & Comments\\
    \midrule
     THE-X~\cite{chen2022x} &-&-&-&-&-&Require  training\\\hline
CipherGPT~\cite{hou2023ciphergpt}  & $\mathbb{R}^{256\times768}$ & 65 & $\textnormal{LAN}^{1}$&  $4.756$ &- &-\\\hline
    SIGMA~\cite{gupta2023sigma} &$\mathbb{R}^{128\times3072}$ &112.64 &$\textnormal{LAN}^{2}$ &0.25 &-&\\\hline
    PrivFormer~\cite{akimoto2023privformer} & -& 1.64  & $\textnormal{LAN}^{3}/\textnormal{WAN}^{4}$& $0.077/10.753$ &- & \\\hline
    Iron~\cite{hao2022iron} & $\mathbb{R}^{128\times768}$ &  871.46 & $\textnormal{LAN}^{1}/\textnormal{WAN}^{5}$ & $16/1158$ &$1.7\times10^{-4}$ & \\\hline
    BOLT~\cite{pang2024bolt} & $\mathbb{R}^{128\times768}$& 599.40 & $\textnormal{LAN}^{1}/\textnormal{WAN}^{5}$ & $14/914$ &- & \\\hline
    NEXUS~\cite{zhang2024secure}& $\mathbb{R}^{128\times768}$ & 0 & $\textnormal{LAN}^{1}/\textnormal{WAN}^{5}$ & $81/81$ & $4.5\times10^{-4}$& Nearly comm. free\\
  \bottomrule
\end{tabular}
\begin{tablenotes}
	\small {
	 \item ${}^{1}$3Gbps, 0.8ms 
            \hspace{0.5em}
            ${}^{2}$9.4Gbps, 0.05ms
            \hspace{0.5em}
            ${}^{3}$4.93Gbps, 1.17ms
            \hspace{0.5em}
            ${}^{4}$97.4Mbps, 141.67ms
            \hspace{0.5em}
            ${}^{5}$100Mbps, 80ms
       }
    \end{tablenotes}
\end{threeparttable}
\end{table}


\section{Reported Evaluation}\label{sec:evaluation}
This section compares the experimental results presented in the selected studies.

\subsection{Models and Datasets}

The transformer models and datasets used for evaluating PTI performance in different studies are significant for comparison. For instance, transformers with more complex structures and a larger number of parameters tend to take longer to do inference, but they often perform better~\cite{jiao2019tinybert}.
Below we will discuss popular models and datasets in PTI studies in detail.

\textbf{Models.} Popular transformer models for PTI studies \cite{li2022mpcformer,chen2022x,hao2022iron,lu2023bumblebee,pang2024bolt,zhang2024secure} mainly include the BERT family (e.g., Bert-Tiny~\cite{turc2019well}, Bert-Base~\cite{devlin2018bert}, Roberta-Base~\cite{liu2019roberta}) and the GPT family (e.g., GPT2-Base~\cite{radford2019language}). Some studies~\cite{dong2023puma,lu2023bumblebee} have also attempted to implement private inference on other transformer models, such as LLaMA-7B~\cite{touvron2023llama} and ViT-Base~\cite{dosovitskiy2020image}. We present a detailed description of various transformer models in Table~\ref{tab:Transformers}.

\begin{table}
\begin{threeparttable}
  \caption{Hyperparameters and Model Sizes of Various Transformer Models}
  \label{tab:Transformers}
  \begin{tabular}{c|c|c|c|c|c}
    \toprule
    Model& Layers (L) & Hidden size (d) & Heads (H)& MLP size & Params\\
    \midrule
    BERT-Tiny~\cite{turc2019well} &  2 & 128 & 2 & 512& 4.4M\\
    BERT-Base~\cite{devlin2018bert} & 12 & 768 & 12 & 3072 & 110M\\
    Roberta-Base~\cite{liu2019roberta} &  12 & 768 & 12& 3072& 125M\\
    BERT-Large~\cite{devlin2018bert} &  24 &1024 &16& 4096& 340M\\\hline
    GPT2-Base~\cite{radford2019language} &  12 &768 &12 & 3072& 117M \\
    GPT2-Medium~\cite{radford2019language} & 24 &1024 &16& 4096& 345M\\
    GPT2-Large~\cite{radford2019language} & 36 & 1280 &16& 5120& 762M\\\hline
    ViT-Base~\cite{dosovitskiy2020image} & 12 & 768 &12 & 3072 & 86M\\\hline
    LLaMA-7B~\cite{touvron2023llama} & 32 &4096 &32 & 11008& 7B\\
  \bottomrule
\end{tabular}
\begin{tablenotes}
	\small {
	 \item \textbf{Note:} MLP means multi-layer perception, i.e., the feed-forward layers.}
    \end{tablenotes}
\end{threeparttable}
\end{table}

BERT's architecture is based on a multi-layer bidirectional Transformer encoder. It uses masked language modeling (MLM) and next sentence prediction (NSP) as pre-training objectives, which allow it to understand the context from both directions (left and right of a token). This bidirectional understanding makes BERT especially effective for tasks requiring contextual inference, such as sentiment analysis, question answering, and language inference.

GPT models are based on a transformer decoder, a left-to-right architecture. They only predict future words based on past and current words, aligning them more closely with traditional language models. The primary training method is causal language modeling (CLM), where the model predicts the next word in a sentence, making GPT highly effective for tasks that involve generating coherent and contextually relevant text.

\textbf{Evaluation Datasets.} To assess the natural language understanding (NLU) abilities of different transformers, we need e general NLU benchmark for evaluation.
Most PTI studies evaluate their performances on the GLUE benchmark~\cite{wang2018glue}, a widely adopted evaluation measure for BERT
and GPT-based transformers. 
GLUE does not pose any constraints on model architecture beyond the ability to process single-sentence and sentence-pair inputs and make predictions, which makes it suitable for evaluating different transformers.
Specifically, the GLUE benchmark contains three different kinds of NLU tasks and nine corresponding corpora: single-sentence tasks (CoLA, SST$-$2), similarity and paraphrase tasks (MRPC, STS$-$B, QQP), and inference tasks (MNLI, QNLI, RTE, WNLI). 

Except for GLUE, some PTI studies~\cite{dong2023puma,lu2023bumblebee,zhang2024secure} also use other benchmarks for evaluations. 
For instance, PUMA~\cite{dong2023puma} and NEXUS~\cite{zhang2024secure} evaluate GPT2 on Wikitext-103 V1~\cite{merity2016pointer} and CBT-CN, respectively. ViT-Base is particularly designed for image processing, and thus BumbleBee~\cite{lu2023bumblebee} uses the ImageNet dataset for its evaluation. 

\begin{table*}[ht]
\centering
\caption{Resource requirements, Tasks performed, Dataset Performance, and Runtime.}
\resizebox{\textwidth}{!}{
\begin{tabular}{l|l|l|l|l|l|lll|l}
\hline
\multirow{2}{*}{Study} & \multirow{2}{*}{Model}     & \multirow{2}{*}{Dataset} & \multirow{2}{*}{I./O. size} & \multirow{2}{*}{Comm.} & \multirow{2}{*}{Comm. Set.} & \multicolumn{3}{l|}{Performance}  & \multirow{2}{*}{Runtime}  \\  \cline{7-9}
&   &   & &   &  & \multicolumn{1}{l|}{Plain} & \multicolumn{1}{l|}{Enc.} & Loss~$\downarrow$ &    \\ \hline

\multirow{4}{*}{MPCFormer~\cite{li2022mpcformer}}  & \multirow{4}{*}{BERT-Base} 
& QNLI & \multirow{4}{*}{(128,1)}  & \multirow{4}{*}{12.089 GB}   & \multirow{4}{*}{(5 Gbps, 1 ms)}
&\multicolumn{1}{l|}{91.7 \%}    & \multicolumn{1}{l|}{90.6 \%}   & 1.1 \%   & \multirow{4}{*}{55.320 s}  \\
&   & MRPC   &  &  &    & \multicolumn{1}{l|}{90.3 \%}     & \multicolumn{1}{l|}{88.7 \%}    & 1.6 \%    &  \\ 
&   & RTE   &  &  &    & \multicolumn{1}{l|}{69.7 \%}     & \multicolumn{1}{l|}{64.9 \%}    & 4.8 \%    & \\
&   & STS-B   &  &  &    & \multicolumn{1}{l|}{89.1 \%}     & \multicolumn{1}{l|}{80.3 \%}    & 8.8 \%    &       \\\hline

\multirow{4}{*}{THE-X~\cite{chen2022x}}  & \multirow{4}{*}{BERT-Tiny} & SST-2 & \multirow{4}{*}{-}  & \multirow{4}{*}{-} & \multirow{4}{*}{-} & \multicolumn{1}{l|}{82.45 \%}    & \multicolumn{1}{l|}{82.11 \%}   & 0.34 \%   & \multirow{4}{*}{-}         \\
&    & MRPC   &    &  &  & \multicolumn{1}{l|}{81.57 \%}     & \multicolumn{1}{l|}{79.94 \%}    & 1.63 \%    & \\ 
&    & RTE   &    &  &  & \multicolumn{1}{l|}{58.56 \%}     & \multicolumn{1}{l|}{58.12 \%}    & 0.44 \%    &  \\
&    & STS-B   &    &  &   & \multicolumn{1}{l|}{72.83 \%}     & \multicolumn{1}{l|}{68.39 \%}    & 4.44 \%    &   \\\hline

\multirow{9}{*}{PUMA~\cite{dong2023puma}}  & \multirow{3}{*}{BERT-Base} & CoLA & \multirow{5}{*}{(128,1)}  & \multirow{3}{*}{10.773 GB} &\multirow{9}{*}{(5 Gbps, 1 ms)}   & \multicolumn{1}{l|}{61.6 \%}    & \multicolumn{1}{l|}{61.3 \%}   & 0.3 \%   & \multirow{3}{*}{33.913 s}         \\
&  & RTE   &    &  &  & \multicolumn{1}{l|}{70.0 \%}     & \multicolumn{1}{l|}{70.0 \%}    & 0.0 \%    & \\ 
&  & QNLI   &    &  &  & \multicolumn{1}{l|}{91.6 \%}     & \multicolumn{1}{l|}{91.6 \%}    & 0.0 \%    &   \\\cline{2-3}\cline{5-5}\cline{7-10}
& Roberta-Base  & -  &    & 11.463 GB &   & \multicolumn{1}{l|}{- }     & \multicolumn{1}{l|}{-}   & -   & 41.641 s  \\\cline{2-3}\cline{5-5}\cline{7-10}
& Bert-Large  & - &    & 27.246 GB &   & \multicolumn{1}{l|}{- }     & \multicolumn{1}{l|}{-}   & -   & 73.720 s  \\\cline{2-5}\cline{7-10}
& GPT2-Base  & \multirow{3}{*}{Wiki.-103}  & \multirow{3}{*}{(32,1)}   & 3.774 GB &   & \multicolumn{1}{l|}{16.284 }     & \multicolumn{1}{l|}{16.284}   & 0.000   & 15.506 s  
\\\cline{2-2}\cline{5-5}\cline{7-10}
& GPT2-Medium  &   &    & 7.059 GB &   & \multicolumn{1}{l|}{12.536 }     & \multicolumn{1}{l|}{12.540}   & -0.004   & 30.272 s  \\\cline{2-2}\cline{5-5}\cline{7-10}
& GPT2-Large  &  &    & 11.952 GB &   & \multicolumn{1}{l|}{10.142}     & \multicolumn{1}{l|}{10.161}   & -0.019   & 54.154 s  \\\cline{2-5}\cline{7-10}
& LLaMA-7B  & - & (8,1)   & 1.794 GB &   & \multicolumn{1}{l|}{-}     & \multicolumn{1}{l|}{-}   & -   & 200.473 s  \\
\hline

\multirow{4}{*}{Iron~\cite{hao2022iron}}  & \multirow{4}{*}{BERT-Base} & SST-2 & \multirow{4}{*}{(128,1)}  & \multirow{4}{*}{280.99 GB}  & \multirow{4}{*}{(3 Gbps, 0.8 ms)}  & \multicolumn{1}{l|}{92.36 \%}    & \multicolumn{1}{l|}{92.77 \%}   & -0.41 \%   & \multirow{4}{*}{475 s}         \\
&   & MRPC   &    &  &   & \multicolumn{1}{l|}{90.00 \%}     & \multicolumn{1}{l|}{89.87 \%}    & 0.13 \%    & \\ 
&   & RTE   &    & &    & \multicolumn{1}{l|}{69.70 \%}     & \multicolumn{1}{l|}{70.76 \%}    & -1.06 \%    &  \\
&    & STS-B   &    &  &   & \multicolumn{1}{l|}{89.62 \%}     & \multicolumn{1}{l|}{89.41 \%}    & 0.21 \%    &  \\\hline

\multirow{7}{*}{BumbleBee~\cite{lu2023bumblebee}}  & \multirow{3}{*}{BERT-Base} & QNLI & \multirow{5}{*}{(128,1)}  & \multirow{3}{*}{-} & \multirow{7}{*}{(1 Gbps, 0.5 ms)}   & \multicolumn{1}{l|}{90.30 \%}    & \multicolumn{1}{l|}{90.20 \%}   & 0.10 \%   & \multirow{3}{*}{-}         \\
&                    & RTE   &    &  &  & \multicolumn{1}{l|}{70.04 \%}     & \multicolumn{1}{l|}{70.04 \%}    & 0.00 \%    &  \\
&                    & CoLA   &    &  &  & \multicolumn{1}{l|}{61.57 \%}     & \multicolumn{1}{l|}{60.82 \%}    & 0.75 \%    &  \\\cline{2-3}\cline{5-5}\cline{7-10}
& BERT-Large  & -  &   & 20.85 GB &  & \multicolumn{1}{l|}{-}     & \multicolumn{1}{l|}{-}   & -  & 404.4 s  \\\cline{2-3}\cline{5-5}\cline{7-10}
& LLaMA-7B  & -  &    & 6.82 GB &  & \multicolumn{1}{l|}{-}     & \multicolumn{1}{l|}{-}   & -   & 832.2 s  \\\cline{2-5}\cline{7-10}
& GPT2-Base  & -  & (32,1)   & 1.94 GB &  & \multicolumn{1}{l|}{-}     & \multicolumn{1}{l|}{-}   & -  & 55.2 s  \\\cline{2-5}\cline{7-10}
& ViT-Base  & ImageNet  & ([224,224,3],1)   & 14.44 GB &  & \multicolumn{1}{l|}{89.44 \%}     & \multicolumn{1}{l|}{89.13 \%}   & 0.31 \%   & 234 s  \\
\hline

\multirow{4}{*}
{BOLT~\cite{pang2024bolt}}  & \multirow{4}{*}{BERT-Base}   & SST-2 & \multirow{4}{*}{-}  & \multirow{4}{*}{25.74 GB} &\multirow{4}{*}{(3 Gbps, 0.8 ms)}  & \multicolumn{1}{l|}{92.36 \%}    & \multicolumn{1}{l|}{92.78 \%}   & -0.42 \%   & \multirow{4}{*}{185 s}          \\
&                   & MRPC   &  &   &    & \multicolumn{1}{l|}{90.00 \%}     & \multicolumn{1}{l|}{89.95 \%}    & 0.05 \%    &  \\ 
&                     & RTE   &  &  &    & \multicolumn{1}{l|}{69.70 \%}     & \multicolumn{1}{l|}{69.31 \%}    & 0.39 \%    &  \\
&                     & STS-B  &  &  &    & \multicolumn{1}{l|}{89.62 \%}     & \multicolumn{1}{l|}{88.44 \%}    & 1.18 \%    &  \\ \hline

\multirow{4}{*}{NEXUS~\cite{zhang2024secure}}  &
\multirow{3}{*}{BERT-Base}  & RTE & \multirow{3}{*}{(128,1)}  & \multirow{3}{*}{0.16 GB} &\multirow{3}{*}{(100 Mbps, 80 ms)}   & \multicolumn{1}{l|}{70.04 \%}    & \multicolumn{1}{l|}{69.88 \%}   & 0.16 \%   & \multirow{3}{*}{1125 s}          \\
&                    & SST-2   & &   &    & \multicolumn{1}{l|}{92.36 \%}     & \multicolumn{1}{l|}{92.11 \%}    & 0.25 \%    &   \\ 
&                    & QNLI   &  &  &    & \multicolumn{1}{l|}{90.30 \%}     & \multicolumn{1}{l|}{89.92 \%}    & 0.38 \%    &  \\\cline{2-10}
& GPT2-Base    & CBT-CN  & -   & -   &-  & \multicolumn{1}{l|}{85.70 \%}     & \multicolumn{1}{l|}{85.31 \%}    & 0.39 \%    &    -                                             \\\hline

\end{tabular}
}
\end{table*}

\section{Conclusion}\label{sec:conclusion}

\bibliographystyle{ACM-Reference-Format}
\bibliography{ref}

\appendix

\section{MPC settings}

We first present the popular MPC settings employed in current secure inference studies. Each of them has different requirements and assumptions on the participating parties. Generally speaking, the 2PC setup is the most natural for secure inference, while 3PC and 2PC-Dealer setups usually support a wider variety of operations and improve the MPC performance. We compare their strengths and weaknesses in Table.~\ref{tab:MPC}.
\begin{itemize}
    \item \textbf{2-party computation (2PC):} This is the basic MPC setting where two participants engage in computation without mutual trust. Each party contributes their input to the computation, but neither can see the other's data. This setting is the most intuitive and commonly used for secure inference, as it only involves two parties and minimizes the complexity of trust and coordination.
    \item \textbf{Honest-majority 3-party computation (3PC):} This configuration introduces an additional "helper" party, increasing the capabilities of the computation. In this model, the adversary may corrupt any single party without compromising the integrity of the computation. The presence of the third party allows for a broader range of operations and enhances the overall performance of MPC.
    \item \textbf{2PC with trusted dealer (2PC-Dealer):} 
    In this scenario, a trusted dealer plays a crucial role during a pre-processing phase by distributing \textit{input-independent} correlated randomness to the two computation parties. This randomness is used to facilitate and speed up computations. The trusted dealer is not involved in the computation itself but enables more efficient processing.
\end{itemize}

\begin{table}[h]
\begin{threeparttable}
  \caption{A Comparison of secure MPC schemes with respect to different key characteristics.}
  \label{tab:MPC}
 \begin{tabular}{c|c|c|c|c|c|c|c}
    \toprule
     \multirow{2}{*}{Scheme} & \multirow{2}{*}{\#Parties } & 
     \multirow{2}{*}{Security Model} &
     \multicolumn{2}{c|}{Overhead} & \multirow{2}{*}{Resilience} & \multirow{2}{*}{Practicality} & \multirow{2}{*}{Scalability}  \\
    \cline{4-5}
    & & & ${\textnormal{Comm.}}^{1}$ & ${\textnormal{Comp.}}^{2}$ & & \\
    \midrule
    2PC &2  &-& + & & &+ &+\\
    3PC &3  &Honest-majority &- &+ & &- &++ \\
    2PC-Dealer &3 &Trusted Dealer & &+ & & &++\\
  \bottomrule
\end{tabular}
\begin{tablenotes}
	\small {
     \item 
     ${}^{1}$ Communication.
     \hspace{0.5em}
     ${}^{2}$ Computation.
     \item  Note that the number of `+' indicates the degree of \textbf{benefit} with respect to a feature, with `+++' denoting the strongest and `+' representing the weakest.
            }
    \end{tablenotes}
\end{threeparttable}
\end{table}

\end{document}